\documentclass[a4paper,11pt]{article}
\usepackage{jheppub}
\usepackage[T1]{fontenc}
\usepackage{makecell}
\usepackage[dvipsnames]{xcolor}
\usepackage{multirow}
\usepackage{xspace}
\usepackage{xcolor} 
\usepackage[colorlinks=true, linkcolor=blue, urlcolor=cyan, citecolor=blue]{hyperref}

\DeclareMathSymbol{\mathbbE}{\mathord}{AMSb}{"45}
\newcommand{\ex}{\mathbbE}
\DeclareMathSymbol{\mathbbH}{\mathord}{AMSb}{"48}
\DeclareMathSymbol{\mathbbR}{\mathord}{AMSb}{"52}

\newcommand{\MG}{\texttt{MadGraph5}\xspace}
\newcommand{\smeft}{\texttt{SMEFTsim}\xspace}

\usepackage{pifont}

\definecolor{myblue}{RGB}{33,74,128}
\definecolor{mygreen}{RGB}{28,120,28}
\definecolor{myorange}{RGB}{200,90,0}
\definecolor{mygray}{gray}{0.95}

\preprint{ANL-201447}

\title{Reusable theory representations for colliders: a demonstrator SMEFT foundation model}

\author[a]{Supratim Das Bakshi,}
\author[a]{T. J. Hobbs,}
\author[a]{Brandon Kriesten}
\affiliation[a]{High Energy Physics Division, Argonne National Laboratory, Lemont, IL, USA}

\emailAdd{sdasbakshi@anl.gov}
\emailAdd{tim@anl.gov}
\emailAdd{bkriesten@anl.gov}

\abstract{
We develop a demonstrator foundation model for collider-scale explorations of the Standard Model Effective Field Theory (SMEFT), constructed from contrastive representations of theoretically simulated neutral-current Drell--Yan cross sections.
Using a controlled sampling of the Warsaw-basis dimension-6 Wilson-coefficient space at $\mathcal{O}(\Lambda^{-2})$, we generate a corpus of high-resolution differential distributions in $m_{\ell\ell}$ and $p_{T}$, augmented by physics-motivated Monte Carlo replicas with correlated uncertainties.
A minimally parameterized encoder network is trained with a supervised contrastive loss to produce a low-dimensional latent manifold on which SMEFT-induced deformations of the Drell--Yan spectrum acquire a well-defined geometric structure. We analyze the resulting embedding and demonstrate that ({\it i}) latent directions correlate with characteristic SMEFT shape distortions, including energy-growing four-fermion contributions and electroweak vertex corrections; ({\it ii}) clusters in the embedding correspond to families of Wilson-coefficient configurations with similar phenomenological impact; and ({\it iii}) the learned representation supports downstream tasks such as classification with uncertainty quantification, anomaly detection, and nearest-neighbor retrieval. While restricted to leading-order SMEFT and simplified uncertainty modeling, this study provides the first step toward a reusable, physics-aligned foundational representation for the theory of New-Physics searches at high-energy colliders.
We outline extensions towards a complete global analyses, including multi-process training corpora, higher-order corrections, and multi-objective pretraining.
}


\begin{document}
\maketitle
\flushbottom


\section{Introduction}
\label{sec:intro}

Modern high-energy experiments at the LHC and related collider facilities operate in a regime of ever-increasing precision. In this setting, simultaneous stress tests of Standard Model (SM) predictions and investigations of New Physics beyond the Standard Model (BSM) through subtle, shape-level deviations in measured kinematical distributions are possible, with many flagship measurements increasingly limited by systematic uncertainties going into the HL-LHC era~\cite{Celada:2024mcf, FCC:2025lpp, terHoeve:2025gey, Maura:2024zxz, Dawson:2022zbb,ILCInternationalDevelopmentTeam:2022izu,Cepeda:2019klc,Boughezal:2022pmb}. As uncertainties come under control, new search opportunities emerge which explore the interplay between SM and BSM signals, {\it e.g.}, in high-mass tails and transverse-momentum spectra where the observed effects of rare processes grow with center-of-mass energy~\cite{Grazzini:2018eyk,Greljo:2021kvv,Angelescu:2020uug,Ethier:2021bye}. A particularly attractive feature of indirect, ``bottom-up,'' BSM approaches is that the investigations remain largely agnostic to the microscopic nature of the particle content of the New Physics which has been introduced. This agnosticism allows for cross-domain searches accessing many observables for physics beyond the direct production channel which may lie outside the reach of current experiments. Effective field theories (EFTs) provide a common language to describe physics at widely separated scales and, in the case of the Standard Model Effective Field Theory (SMEFT), connecting the effects of a specific ultraviolet UV-complete theories to lower-energy effective interactions which produce deviations from SM predictions. These properties of SMEFT, alongside recent advances in AI/ML, motivate the chief aim of this study: the development of a demonstrator foundation model which learns a universal representation of collider distributions, transferable across observables and processes, with the ability to encode the subtleties of SMEFT as reflected by the collider signatures it produces.

SMEFT~\cite{Grzadkowski:2010es,Brivio:2017vri,Isidori:2023pyp} has now been extensively developed over the past $\sim$decade, enjoying widespread adoption in support of BSM searches, especially at high-energy colliders. SMEFT augments the SM Lagrangian by introducing higher-dimensional operators, $d\!>\!4$, which preserve the standard gauge symmetries, $\mathrm{SU}(3)_{C} \times \mathrm{SU}(2)_{L} \times \mathrm{U}(1)_{Y}$, providing a model-agnostic way of introducing New-Physics signatures in collider observables measurable at facilities like the LHC. Standard power-counting rules \cite{Assi:2025zmp} for the partonic cross section admits a tower of higher-dimensional operators through an expansion in inverse powers of the high-energy cutoff $1/\Lambda$:
\begin{eqnarray}
    \sigma_\mathrm{SMEFT} &=& |\mathcal{M}_\mathrm{SM}\ +\ \frac{1}{\Lambda^{2}}\mathcal{M}_{d6}\ +\ \frac{1}{\Lambda^{4}}\mathcal{M}_{d8}\ +\ \dots|^{2} \nonumber \\
    &=& |\mathcal{M}_\mathrm{SM}|^{2}\Bigg( 1 + \frac{2\Re e}{\Lambda^{2}}\frac{\mathcal{M}_\mathrm{SM}^{*}\mathcal{M}_{d6}}{|\mathcal{M}_\mathrm{SM}|^{2}} \nonumber \\
    && \qquad\qquad\quad+ \frac{1}{\Lambda^{4}}\frac{|\mathcal{M}_{d6}|^{2}}{|\mathcal{M}_\mathrm{SM}|^{2}} + \frac{2\Re e}{\Lambda^{4}}\frac{\mathcal{M}_\mathrm{SM}^{*}\mathcal{M}_{d8}}{|\mathcal{M}_\mathrm{SM}|^{2}} + \dots\Bigg)\, .
\end{eqnarray}
In this work, we truncate the expansion at the linear, dimension-6 interference contribution and, to ensure EFT validity, we investigate observables such that $\hat{s}/\Lambda^{2}$ is controllable. An EFT matching can then be performed ``top-down'' to relate UV-complete model parameters ({\it e.g.}, an introduction of a new $\mathrm{U}(1)$ gauge symmetry such as a heavy $Z'$ \cite{Dawson:2024ozw}, an extended $\mathrm{SU}(2)$ gauge symmetry for a heavy $W'$, or even extensions of the Higgs sector with added multiplets such as in the 2HDM \cite{DasBakshi:2024krs,Dawson:2022cmu}) to SMEFT Wilson coefficients, $C_{i}^{(d)}(\mu)$, at an energy scale $\mu$, followed by renormalization group (RG) running to the experimental scale --- effectively creating a dictionary to translate between the two \cite{deBlas:2017xtg,Gherardi:2020det,Anisha:2021hgc,Guedes:2024vuf}.

One of the central aims of SMEFT phenomenology is to thoroughly constrain the Wilson coefficient parameter space through global analyses frameworks, successively traversing  higher dimensional operators and fitting as many of the corresponding Wilson coefficients as possible within the context of a particular (or set of) collider processes. In recent years, a number of global analysis efforts have attempted simultaneous fits of SM physics (such as QCD hadronic structure in PDFs and electroweak couplings) alongside effective field theory coefficients; where the goal is to address degeneracies and potential contaminations when SM-only and BSM+SM hypotheses are fitted separately. Sensitivity to New Physics in differential cross sections is often derived from shape-level information on, {\it e.g.}, the high-invariant-mass tails of charged- and neutral-current Drell--Yan in $pp$ collisions, and by top-quark and jet observables, where energy growth enhances four-fermion contact interactions. 

A common challenge in global analyses of SMEFT Wilson coefficients is the vast dimensionality of the available parameter space. Beyond dimension-4, the full flavor (baryon-number conserving) Warsaw basis contains 2499 independent Wilson coefficients at dimension-6 and exponentially increases to 44,807 Wilson coefficients at dimension-8. Taking just the number of independent operators (disregarding flavor indices) reduces the dimensionality of the parameter space to 59 total operators for dimension-6~\cite{Grzadkowski:2010es} and 175 bosonic operators (not including four-fermion operators) for dimension-8 \cite{Murphy:2020rsh,Ren:2022tvi}. Phenomenological studies often truncate the operator basis still further, concentrating on operator classes which dominate the interaction strength or are weakly constrained, in contrast to those which are tightly constrained ({\it e.g.}, operators shifting precisely known SM parameters or chirality-flipping dipole-like structures). Further complications arise when truncating the EFT expansion by introducing consistency choices in $\mathcal{O}(\Lambda)$. If the series expansion is continued to $\mathcal{O}(\Lambda^{-4})$, interactions between dimension-6 squared terms versus the linear dimension-8 contributions must both be taken into account. The problem of traversing this large dimensional parameter space, and parsing the underlying correlations to understand the full SMEFT theory space, is an enormous task which strains current EFT global analyses frameworks \cite{Boughezal:2020uwq,Allwicher:2022gkm}.

Modern computing tools have always evolved hand-in-hand with theoretical particle physics discoveries. As theory and experiment enter a precision era with the upcoming HL-LHC, the use of novel computing tools must also mature. Currently, AI/ML assisted calculations have interrogated many areas of theory through models built for particular tasks; however, their narrow focus on particular problems do not allow for extrapolation to physics discovery beyond their training context. This manuscript builds upon this background to develop a demonstrator foundation model for SMEFT signatures in collider physics, with the aim of investigating the theory at a fundamental level beyond global fitting. Foundation models can be defined as broadly pretrained shared embedding spaces which hold data in a variety of modalities to uncover subtle, or even hidden, correlations between model inputs. In the context of HEP theory, the embedding space is a shared mathematical manifold on which data can be dimensionally reduced in a nonlinear fashion. The data modalities in this context represent the ``language'' through which we can interrogate the theory which, in the case of SMEFT, is the collider observables. The pretraining tasks are any learning objective which helps the foundation model encode correlations and encapsulate the internal properties of the SMEFT predictions, such as contrastive separation among distinct SMEFT scenarios, event generation, SM/BSM classification, positivity predictions for Wilson coefficients, {\it etc}. These foundational embeddings are then used for a variety of downstream tasks such as classification, anomaly detection, and uncertainty quantification through fine-tuned head models. 

Global analyses of SMEFT Wilson coefficients typically proceed along a well-defined pipeline: select a dataset or a group of datasets which are specifically chosen due to sensitivity to particular BSM scenarios of interest in a defined kinematic setting; specify the theory set-up, such as EFT truncation and validity regions, PDF sets, the perturbative order of the QCD or electroweak calculations, and SM input parameters; choose a set of tractable Wilson coefficients which maximally deform potential shapes of kinematical distributions; and perform a likelihood fit over the chosen Wilson coefficients to extract bounds and establish potential correlations~\cite{Ellis:2018gqa,Dawson:2020oco,Corbett:2021eux,Giani:2023gfq}. Recent studies have begun to unfreeze the parameters of the PDFs used in the EFT global fit, and investigated potential BSM-PDF correlations or spurious absorption of BSM signatures into the hadronic structure --- even so much as identifying potential measurements which are least sensitive to BSM physics  \cite{Greljo:2021kvv,Bissolotti:2023vdw,Hammou:2023heg}. In this context, a SMEFT foundation model is complementary to standard global analyses chains. The purpose of a foundation model is to develop a shared embedding space in which kinematical distributions are correlated across processes, energies, and analysis choices in an effort to learn the underlying theory through the language of the model inputs (in this case, the collider physics observables). The trained representation supports tasks such as inference/regression of operator combinations, nearest-neighbor retrieval (which SMEFT scenario does a particular distribution look like?), Out-of-distribution (OOD)/anomaly detection, uncertainty quantification, and potential generation of distributions when paired with a decoder. This emphasizes that, complementary to a SMEFT global analysis, inference is not the end goal of a foundation model, but rather its reusable embedding structure. This puts foundation models on another level beyond traditional fitting. The two paradigms are connected naturally, which also allows for potential cross-domain inference and interactions with agentic systems~\cite{Bakshi:2025fgx}.

\begin{figure}
    \centering
    \includegraphics[width=0.75\linewidth]{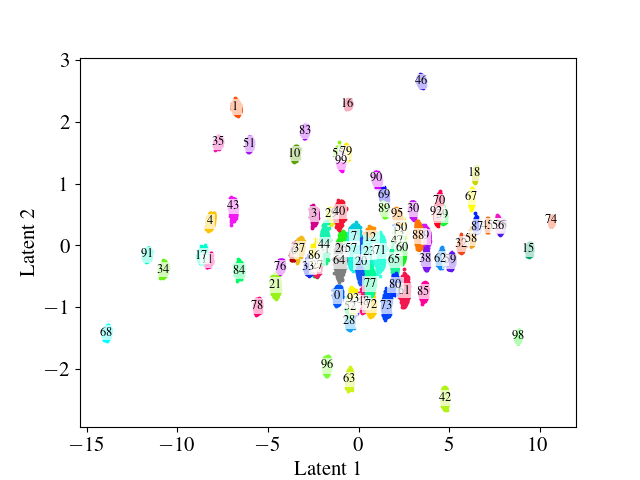}
    \caption{Example foundational embeddings of the differential distributions of the invariant mass and the transverse momenta in two dimensions with each SMEFT scenario labeled.}
    \label{fig:placeholder1}
\end{figure}

In this work we show progress towards a SMEFT foundation model by developing and interrogating a physics-aware encoder model which has been trained on theoretical differential spectra of the dilepton invariant mass and transverse momentum for the neutral-current Drell--Yan process, which we take as a fundamental demonstration case. Consistent with the foundational nature of the model, embeddings are then used to demonstrate several downstream tasks; in this study, these include nearest-neighbor retrieval, classification with uncertainty quantification, and OOD/anomaly detection. We generate $\mathcal{O}(10^2)$ samplings of the SMEFT parameter space (effectively, we refer to these below as distinct collider ``universes'' with respect to their non-standard interactions) by scanning relevant dimension-6, baryon-number conserving independent SMEFT Wilson coefficients [assuming the Warsaw basis and approximate $\mathrm{U}(3)^{5}$ flavor symmetry]; we specifically admit those operators which give a non-zero contribution to the total inclusive cross section at the order of $\mathcal{O}(\Lambda^{-2})$, varying the associated Wilson coefficients over phenomenologically relevant ranges, and fix the SMEFT truncation to the linear interference term in regions where the EFT validity is still controlled at $\hat{s}/\Lambda^{2}$. These choices reduce flavor degeneracies and yield a tractable, reproducible pretraining corpus which we can relax in staged extensions. The differential distributions of the hard-scattering interaction are generated through simulations in \MG \cite{Alwall:2011uj}, taking the CT18LO~\cite{Yan:2022pzl} parton densities as our fixed PDF set. We take the leading-order PDFs to match the perturbative order of the SMEFT and QCD expansions within our simulated events; the distributions have uncertainties which are statistically modeled to build a robust representation that is sensitive to the SMEFT deformations. The essential physics questions we probe are: 
\begin{itemize}
    \item can the latent embedding disentangle the SMEFT scenarios according to their corresponding observable shape deformations?
    \item do latent directions align with SMEFT operator classes as they appear phenomenologically?
    \item does the latent representation allow for the identification of OOD/anomalous distributions?
\end{itemize}
We pre-train the encoder model with a contrastive learning objective which effectively pushes different SMEFT scenarios apart and pulls similar distributions closer together, yielding a semantic embedding space where trajectories in the embeddings correspond to controlled changes in the SMEFT Wilson coefficient structure. The frozen foundational embedding is then used to perform classification with uncertainty quantification, anomaly detection, and nearest-neighbor identification.

The foundational embedding representation which we have introduced here is a first step towards a reusable foundation model for SMEFT phenomenological studies. Future work includes: ({\it i}) scaling the pretraining corpus beyond neutral-current Drell--Yan to multi-process, multi-energy datasets including rapidity, angular observables, and double differential distributions; ({\it ii}) incorporating higher-order/higher-dimensional contributions under controlled power counting ({\it e.g.}, implementing NLO QCD and higher-loop SMEFT contributions, $\mathcal{O}(\Lambda^{-4})$ terms which include dimension-6 squared as well as linear dimension-8, RG evolution and potential matching); ({\it iii}) investigating the inclusion of simulation elements beyond the hard-scattering interaction, including parton-shower and detector effects; ({\it iv}) and exploring transformer-style architectures with physics-aware tokenization and multi-objective pretraining tasks which balance contrastive learning with predictive tasks. Additionally, it is important to broaden the model's exposure to SM parameter shifts such as changes in the strong coupling, and the PDF uncertainties. Downstream, we aim to extend the encoder with event generators, Wilson coefficient inference, as well as BSM+SMEFT matching. The foundation model is an extensible framework which also allows future interactions with SMEFT global analyses or agentic systems. Collectively, these directions establish a practical route to a SMEFT foundation model that complements global analyses and unlocks scalable, transferable reasoning about collider data.

The remainder of this manuscript is as follows: in Sec.~\ref{sec:smeft} we introduce the SMEFT formalism core to our calculations in two subsections, with Sec.~\ref{sec:theory} serving as a theory primer for those less familiar with the fundamentals of SMEFT, and Sec.~\ref{sec:collider_pheno} describing the relevant collider phenomenology and how SMEFT affects observables relevant to this study; Sec.~\ref{sec:global_analyses} describes current efforts in global SMEFT analyses. We similarly describe the foundation model underpinnings in Sec.~\ref{sec:foundation_models}, including Sec.~\ref{sec:overview}, which gives a general description of foundation models and how they work, Sec.~\ref{sec:data}, which describes our choices for generating data as well as a description of the model architecture and pretraining task(s) in Sec.~\ref{sec:arch}; and the available downstream tasks in Sec.~\ref{sec:downstream}. We then show the results of our model in Sec.~\ref{sec:results} and finally conclude with an outlook toward future work in Sec.~\ref{sec:conclusions}.
 
\section{Standard Model Effective Field Theory}
\label{sec:smeft}

In this section, we summarize the theoretical framework and conventions that we follow in our analysis. We briefly outline the effective field theory approach adopted throughout this work and fix the notation and assumptions that will be used in the remainder of the paper.

\subsection{SMEFT Theory Basics and Notation}
\label{sec:theory}

	The SMEFT Lagrangian extends the SM by an infinite tower of	$\mathrm{SU}(3)_C\! \times\! \mathrm{SU}(2)_L\! \times\! \mathrm{U(1)}_Y$ gauge–invariant operators constructed from the SM fields and respecting its symmetries.
    In this work, we truncate the expansion at dimension-6 and assume all Wilson coefficients to be real. We adopt the Warsaw basis~\cite{Buchmuller:1985jz,Grzadkowski:2010es} and express the Lagrangian as,
	\begin{align}
		\mathcal{L}_{\text{SMEFT}}
		= \mathcal{L}_{\text{SM}}
		+ \frac{1}{\Lambda^{2}} \sum_{i} C_{i}\, O_{i},
		\label{eq:smeftlageq1}
	\end{align}
	where $O_i$ denote the complete and independent set of dimension-6 operators, and $C_i$ are the corresponding Wilson coefficients containing the effects of heavy physics at the scale $\Lambda$.

    The Warsaw basis contains 2499 independent Wilson coefficients when restricting to baryon- and lepton-number conserving dimension-6 operators \cite{Alonso:2013hga}.
    We impose a $\mathrm{U}(3)^5$ flavor symmetry on the Wilson coefficient space, under which each fermion multiplet transforms as
    \begin{align}        
    q_i \rightarrow (U^q)_{ij} q_j, \qquad
    \ell_i \rightarrow (U^\ell)_{ij} \ell_j, \qquad
    u_i \rightarrow (U^u)_{ij} u_j, \qquad
    d_i \rightarrow (U^d)_{ij} d_j, \qquad
    e_i \rightarrow (U^e)_{ij} e_j,
    \end{align}
    where $U^\psi$ ($\psi = q,\ell,u,d,e$) are unitary matrices in the corresponding $\mathrm{U}(3)$ flavor groups.
    Each fermion multiplet transforms in the fundamental representation of its associated flavor symmetry.
    Imposing $\mathrm{U}(3)^5$ symmetry reduces the number of independent Wilson coefficients to 70.
    Among these, the operators $O_{ll}$, $O_{qq}^{(1)}$, $O_{qq}^{(3)}$, $O_{uu}$,  $O_{dd}$, $O_{quqd}^{(1)}$, and $O_{quqd}^{(8)}$ contain two independent flavor contractions and thus yield two independent coefficients each (see Sec.~3.2 in \cite{Brivio:2020onw}).
    In this work, we further restrict to real-valued Wilson coefficients, focusing on CP-conserving New-Physics effects, leaving a total of 60 Wilson coefficients in our implementation of the Warsaw basis; we note that the assumption of CP conservation might in principle be relaxed, and we leave to future work more flexible SMEFT parametrizations without loss of generality for the calculation presented here.

    We work at tree-level in the SMEFT and retain only terms up to linear order in the dimension-6 SMEFT Wilson coefficients.
    The electroweak sector is parameterized in terms of the set of input parameters, $\{\hat{m}_W,\, \hat{m}_Z,\, \hat{G}_F\}$.
    In SMEFT, four-fermion operators contribute to muon decay, modifying the relation between the electroweak (EW) vacuum expectation value (vev) and the Fermi constant.
    In particular~\cite{Brivio:2017btx},
    \begin{align}
    	\hat{v}_T &= \frac{1}{2^{1/4}\sqrt{\hat{G}_F}}; 
    	\qquad
    	\delta G_F 
    	= \frac{1}{\hat{G}_F}
    	\left( C_{Hl}^{(3)} - \frac{C_{ll}+C_{ll}'}{4} \right),
    	\label{eq:cllingf}
    \end{align}
    so that the relation between the SMEFT vev and the electroweak input parameter $\hat{G}_F$ receives corrections controlled by the flavor-universal combinations of the four-fermion Wilson coefficients shown above and as typically parametrized in the Warsaw basis.
    
\subsection{SMEFT Collider Phenomenology}
\label{sec:collider_pheno}

    Drell–Yan production provides a theoretically clean and experimentally precise probe of contact interactions in the SMEFT \cite{Dawson:2018dxp}.
    The SM prediction is known to high perturbative accuracy enabling precision tests. In the SMEFT, the interference of quark-lepton four-fermion operators with the SM grows with the partonic energy, rendering the high-mass dilepton tails especially sensitive to $\mathcal{O}(\Lambda^{-2})$ effects~\cite{Greljo:2017vvb}.
    These features motivate dedicated SMEFT analyses of Drell--Yan.
    Complementarity with electroweak precision observables (EWPOs) is well-established.
    EWPOs tightly constrain $Z$–pole couplings and Drell--Yan extends sensitivity to energy–growing directions and flavor structures less accessible at LEP --- necessitating consistent global treatments in SMEFT~\cite{Berthier:2015oma,Corbett:2025oqk}.
    
    Recent computations further include mixed QCD–electroweak SMEFT corrections for neutral-current Drell--Yan, supporting the
    use of linearized SMEFT predictions in LHC precision fits~\cite{Dawson:2021ofa,Dawson:2018dxp,ElFaham:2024egs}.
    Finally, the EFT-truncation and validity domain in the Drell--Yan tails have been systematically studied, providing practical prescriptions for reporting limits differentially in the invariant mass (or partonic energy) and assessing the impact of potential dimension-8 effects~\cite{Contino:2016jqw,Boughezal:2021tih,Allwicher:2024mzw}.
    
    \begin{figure}
    \centering
    \includegraphics[width=0.5\linewidth]{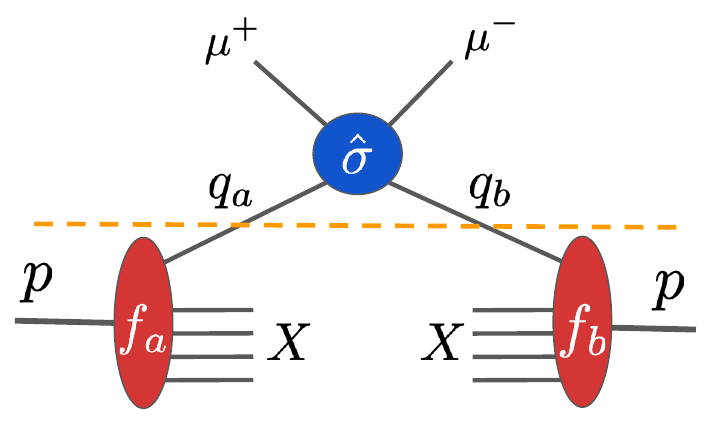}
    \caption{An illustration of the neutral-current Drell--Yan process, $p\,p \rightarrow \mu^{+}\,\mu^{-} + X$, indicating the PDF contributions quantified by $f_{a,b}$, corresponding to the partons $\{q_a,q_b\} \in [q, \bar{q}, g]$, and the hard-scattering kernel, $\hat{\sigma}$. The orange line demonstrates a factorization of the hard scatter from the nonperturbative matrix elements containing the PDFs.}
    \label{fig:placeholder2}
    \end{figure}

    The contributions of dimension-6 SMEFT operators to Drell--Yan production have been computed in the literature \cite{Alioli:2018ljm,Dawson:2018dxp,Greljo:2021kvv,Dawson:2021ofa,Boughezal:2022nof,Allwicher:2022gkm,Allwicher:2024mzw,Hiller:2025hpf,Corbett:2025oqk}.
    In this work, we focus on the neutral-current channel, $p\, p \rightarrow \mu^{+} \,\mu^{-}$, which provides a clean and well-measured probe of quark-lepton contact interactions and electroweak vertex corrections. The subset of Warsaw-basis operators that contribute to this process at tree level is summarized in
    Table~\ref{tab:processWCs},
    and represent thirty-one interactions arising from different SMEFT classes as relevant for the neutral-current Drell--Yan amplitude~\cite{Dedes:2017zog}.
    Among these, the bosonic operators $O_{\varphi \Box}$, $O_{\varphi D}$, and $O_{\varphi WB}$ modify the electroweak symmetry-breaking structure and hence induce shifts in the masses
    and kinetic mixings of the $Z$-boson and the Higgs. The operator $O_{\varphi G}$ affects the Higgs-gluon effective coupling.
    The Yukawa-like operators $O_{e\varphi}$, $O_{u\varphi}$, and $O_{d\varphi}$ generate corrections to the fermion masses and Yukawa interactions.
    The dipole operators $O_{eW}$, $O_{eB}$, $O_{uW}$, $O_{uB}$, $O_{dW}$, and $O_{dB}$ induce chirality-flipping couplings between fermions and the electroweak gauge bosons.
    The current-type operators $O_{\varphi \ell}^{(1)}$, $O_{\varphi \ell}^{(3)}$, $O_{\varphi e}$, $O_{\varphi q}^{(1)}$, $O_{\varphi q}^{(3)}$, $O_{\varphi u}$, and $O_{\varphi d}$ modify the neutral-current interactions between the $Z$-boson and fermions, leading to vertex corrections measurable in dilepton production.
    Finally, the four-fermion operators provide contact-interaction contributions to the amplitude, whose interference with the SM dominates in the high-$p_T$ (or high-invariant-mass) region of the dilepton differential cross sections.

\begin{table}[ht!]
    \centering
    \renewcommand{\arraystretch}{2}
    \begin{tabular}{|c|c|c||c|c|c|}
        \hline
        \multirow{2}{*}{Process} & \multicolumn{5}{c|}{Operators} \\ \cline{2-6}
         & $\mathcal{O}(\hat{s}/\Lambda^{2})$ & $\mathcal{O}(v^{2}/\Lambda^{2})$ & $\mathcal{O}(v^{4}/\Lambda^{4})$ & $\mathcal{O}(v^{2}\hat{s}/\Lambda^{4})$ & $\mathcal{O}(\hat{s}^{2}/\Lambda^{4})$ \\ \hline
        $ p p \rightarrow \mu^+ \mu^-$ & \thead{$O_{\ell q}^{(1)}, O_{\ell q}^{(3)}$\\ $O_{eu}, O_{ed}$ \\ $O_{\ell u}, O_{\ell d}, O_{qe}$ }& \thead{$O_{\varphi \ell}^{(1)}, O_{\varphi \ell}^{(3)}, O_{\varphi e}$ \\$ O_{\varphi q}^{(1)}, O_{\varphi q}^{(3)}, O_{\varphi u}$\\$ O_{\varphi d}, O_{\varphi D}$ \\$O_{\varphi WB}, O_{\ell \ell}^{(1)}$} & \thead{$O_{\varphi \Box}$\\ $O_{e\varphi} $ \\ $ O_{u\varphi}$ \\ $O_{d\varphi}$} & \thead{$O_{eW},O_{eB}$ \\$  O_{uW}, O_{uB}$ \\ $ O_{dW}, O_{dB}$\\$O_{\varphi G}$} & \thead{$O_{\ell e q d}$\\$ O_{\ell e q u}^{(1)}$\\ $O_{\ell e q u}^{(3)}$ } \\ \hline
    \end{tabular}
    \caption{Warsaw-basis dimension-6 operators contributing to neutral-current Drell--Yan production $pp \rightarrow \mu^{+}\mu^{-}$ at tree level.
    The operators are grouped according to the leading power counting of their interference with the SM amplitude or squared contributions: energy-enhanced four-fermion contact interactions scaling as $\mathcal{O}(\hat{s}/\Lambda^{2})$; vertex and gauge-structure corrections entering at
    $\mathcal{O}(v^{2}/\Lambda^{2})$ ($O_{ll}^{(1)}$ enters through the input parameter $G_F$); Yukawa-type two-fermionic operators contributing as $\mathcal{O}(v^{4}/\Lambda^{4})$; dipole  operators and $O_{\varphi G}$ contributing as $\mathcal{O}(v^{2}\hat{s}/\Lambda^{4})$; and scalar semi-leptonic operators scale as $\mathcal{O}(\hat{s}^{2}/\Lambda^{4})$ in the limit of massless SM
    fermions.
    Here $v$ denotes the electroweak scale and $\hat{s}$ is the partonic center-of-mass energy. We compared the scaling of these operators with the results in Ref.~\cite{Boughezal:2021tih} and found them to be in agreement.}
    \label{tab:processWCs}
\end{table}

    \begin{figure}
    \centering
    \includegraphics[width=0.46\linewidth]{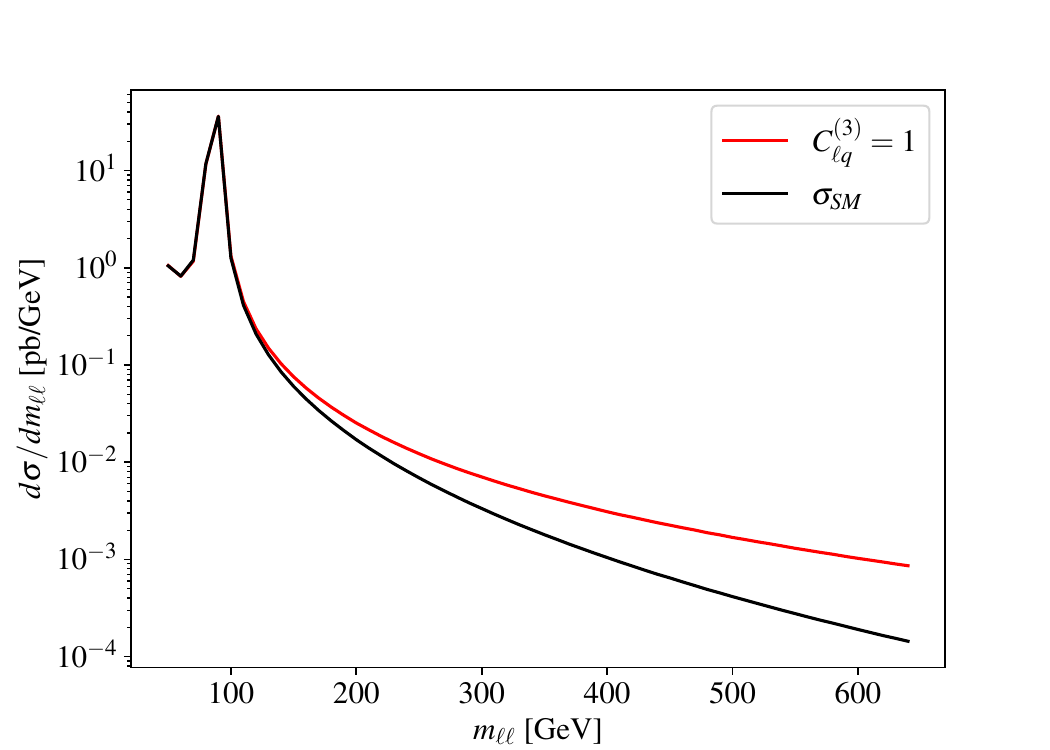}
    \includegraphics[width=0.435\linewidth]{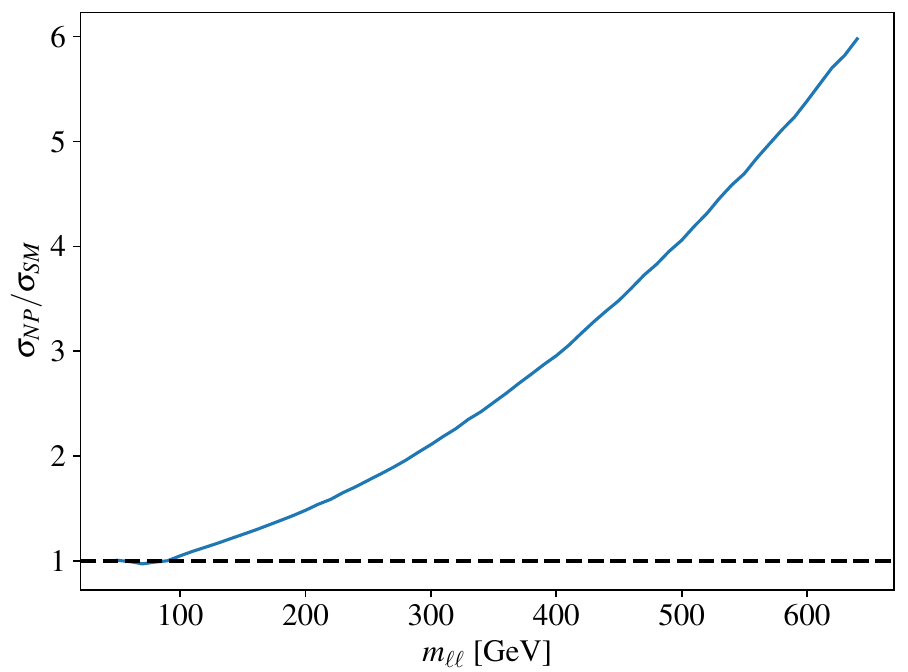}
    \includegraphics[width=0.46\linewidth]{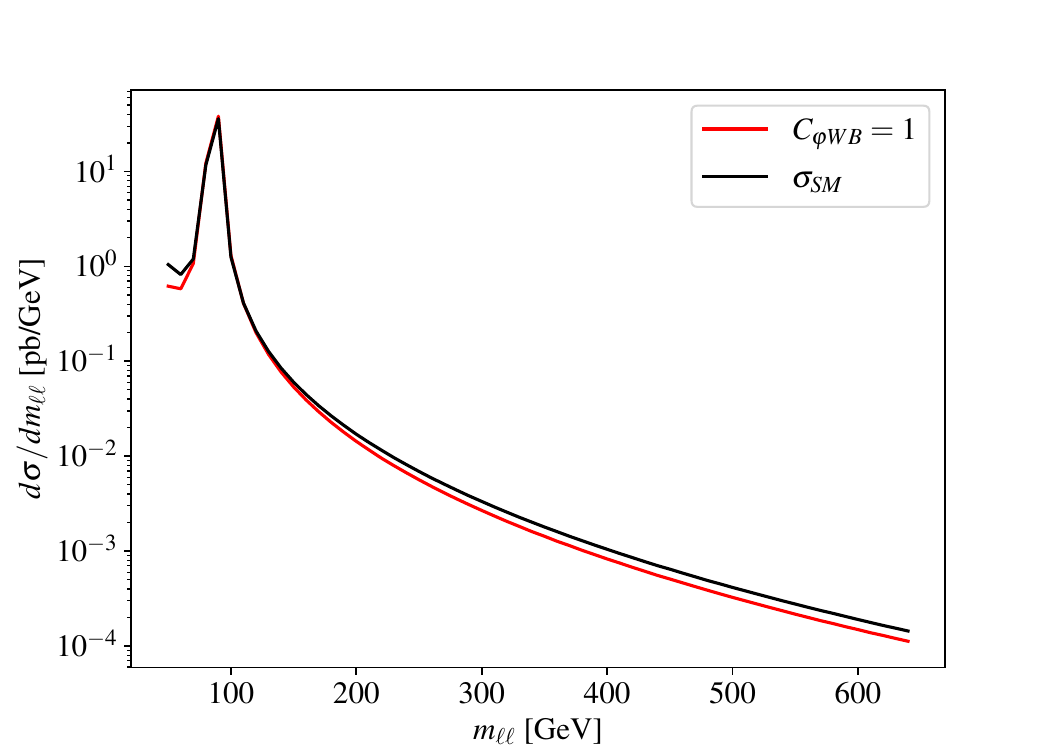}
    \includegraphics[width=0.45\linewidth]{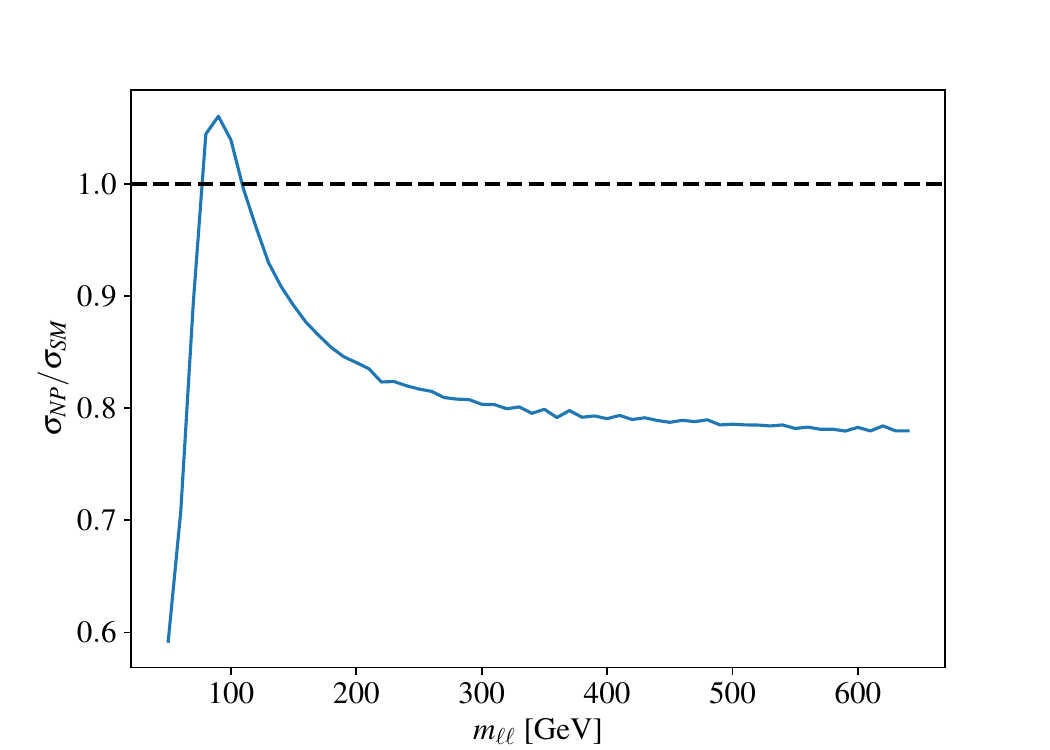}
    \caption{Two examples of typical SMEFT contributions to the neutral-current Drell--Yan process where ({\it top}) is an insertion of the four-fermion SMEFT operator $\mathcal{O}_{\ell q}^{(3)}$ with Wilson coefficient equal to one and the corresponding ratio with the SM, notice that the deviation from this insertion grows as a function of energy coinciding with Table \ref{tab:processWCs}. ({\it bottom}) is an insertion of the bosonic operator $\mathcal{O}_{\varphi WB}$ with Wilson coefficient equal to one and its ratio with the SM, notice that the ratio peaks around the $Z$-pole and remains constant at high masses.}
    \label{fig:placeholder3}
    \end{figure}

\subsection{Global Analysis}
\label{sec:global_analyses}

The neutral-current Drell--Yan process, $ p \, p \to \mu^{+} \mu^{-}$, plays an important role in constraining SMEFT, owing to its clean final state, high statistics, and well-understood electroweak structure.
At dilepton invariant masses well above the $Z$-pole, the differential spectrum $d\sigma/dm_{\ell\ell}$ becomes dominantly sensitive to semileptonic four-fermion structures of the form $(\bar{q}\gamma_{\mu} q)(\bar{\ell}\gamma^{\mu} \ell)$, which interfere with the SM amplitude at $\mathcal{O}(\Lambda^{-2})$ and contribute squared terms at $\mathcal{O}(\Lambda^{-4})$~\cite{Panico:2021vav,Boughezal:2022nof}.
Because the partonic subprocess $q\bar{q}\to \mu^+\mu^-$ factorizes cleanly and involves no QCD interactions at leading order, Drell--Yan provides a clean process in which the scaling behavior of SMEFT deformations can be directly mapped onto the high-$m_{\ell\ell}$ phase space.
Consequently, Drell--Yan constrains directions in Wilson-coefficient space that remain weakly bounded by electroweak precision observables or low-energy scattering~\cite{Isidori:2023pyp}. 
Moreover, the sensitivity of Drell--Yan to renormalization-group (RG)--induced mixing among semileptonic operators~\cite{Boughezal:2024zqa} enables a consistent connection between $Z$-pole measurements and the multi-TeV regime probed at the LHC, ensuring that SMEFT inferences remain valid across disparate energy scales. 
Modern global fits therefore incorporate high-mass Drell--Yan measurements from ATLAS and CMS as a 
statistically powerful probe of non-resonant New Physics in the quark-lepton sector, 
especially in the context of HL-LHC precision targets~\cite{Hiller:2025hpf}.

Despite its central role in SMEFT analyses, the extraction of robust constraints from the neutral-current Drell--Yan process is complicated by several intertwined theoretical and statistical challenges. First, Drell--Yan observables probe only a restricted set of linear combinations of four-fermion Wilson coefficients, leading to well-known degeneracies and approximate flat directions in the global likelihood functions typically used in EFT fits, particularly among operators that generate identical distortions of the high-$m_{\ell\ell}$ spectrum or share similar quark-flavor structure~\cite{Boughezal:2020uwq}.
These degeneracies are further entangled with RG evolution, which mixes semileptonic operators between the electroweak scale and multi-TeV energies; consequently, constraints derived solely at the measurement scale neglect nontrivial correlations induced by SMEFT running~\cite{Bartocci:2024fmm}.
A second source of complexity arises from the truncation of the SMEFT expansion: analyses performed at $\mathcal{O}(\Lambda^{-2})$ retain only interference terms, while those incorporating $\mathcal{O}(\Lambda^{-4})$ capture squared contributions that dominate in the high-energy tail.
The choice of truncation scheme induces systematic shifts in the inferred Wilson-coefficient posteriors~\cite{Boughezal:2021tih}, requiring careful treatment to maintain theoretical consistency.
Finally, modern Drell--Yan datasets involve multi-dimensional binning and correlated systematic uncertainties, which render likelihood construction computationally demanding; naive factorization of uncertainties or Gaussian approximations fail to capture the full structure of the experimental covariance, especially in HL-LHC projections. 
These combined effects underscore the need for methodologies capable of encoding high-dimensional correlations, RG-induced structure, and truncation-dependent behavior in a unified statistical framework.

The limitations of traditional SMEFT global-fit methodologies motivate the development of foundation models that learn statistically faithful latent representations of empirical observables like those associated with the Drell--Yan process.
In an autoencoder-type construction, the high-dimensional  theoretical predictions, $\vec{t}(\vec{C}) \in \mathbb{R}^{N}$, can be mapped to a compressed  latent space, $z \in \mathbb{R}^{k}$, with $k \ll N$, through a non-linear, information-preserving encoder. Here, $\vec{C}\! \in\! \{C_i\}$ represents a unique array of Wilson coefficients fully specifying the SMEFT contributions to a given set of Drell--Yan observables.\footnote{In the analysis of the neutral-current Drell--Yan process presented in this article, \(C_i\) denote the Wilson coefficients corresponding to the operators listed in Table~\ref{tab:processWCs}.}
A corresponding decoder then maps this latent representation back to a set of reconstructed predictions, $\tilde{\vec{t}}(z)$, for the family of
Drell--Yan observables; in this manner, the latent manifold captures the full truncated SMEFT theory --- namely, the dependence of
Drell--Yan observables to variations in the Wilson coefficients.
Crucially, the latent space learns the \emph{correlated} structure of SMEFT-induced deformations in the Drell--Yan cross section: bins whose distortions are linked by PDF systematics, scale variations, or operator mixing become coherent directions in $z$, while statistically independent components are separated into orthogonal latent modes.
This provides a compact representation that preserves the Fisher information and the full experimental covariance structure~\cite{hintonandSalakhutdinov,Zacherl_2021}. As a result, the global likelihood may be constructed in latent space,
\begin{align}
\mathcal{L}(\vec{C}) \;\propto\; 
\exp\!\left[-\tfrac{1}{2}\big(z(\vec{C})-z_{\rm data}\big)^{T}
\Sigma_{z}^{-1}
\big(z(\vec{C})-z_{\rm data}\big)\right],
\end{align}
where $\Sigma_{z}$ is the covariance projected onto the learned manifold.
This approach naturally resolves degeneracies between Wilson-coefficient directions that produce near-identical shapes in the original observable space, while faithfully encoding non-trivial correlations among bins that are typically lost in factorized or Gaussian-approximated treatments.
The latent-space formulation therefore offers a statistically robust foundation on which Drell--Yan constraints can be integrated into global SMEFT analyses without sacrificing the high-dimensional structure inherent in real experimental data.

Beyond its statistical advantages, the latent representation learned by a foundation model provides a natural interface with the theoretical structure of the SMEFT.
Since the Drell--Yan amplitude depends linearly on Wilson coefficients of dimension-6 operators at $\mathcal{O}(\Lambda^{-2})$, with quadratic dependence entering at $\mathcal{O}(\Lambda^{-4})$, the manifold of predicted spectra, $\{\vec{t}(\vec{C})\}$, forms a low-dimensional, smoothly curved surface in observable space, spanned by a finite set of physically meaningful directions corresponding to operator-induced deformations of the partonic amplitude. We emphasize that we restrict our present analysis to dimension-6 only, and do not consider dimension-8 operators which enter linearly through SM interference at $\mathcal{O}(\Lambda^{-4})$.
The encoder learns this structure directly: latent coordinates, $z_{a}$, generally align with combinations of operators that produce distinct variations in the shapes of differential cross sections, while degeneracies associated with, {\it e.g.}, PDF variations independent of SMEFT parameters tend to collapse collinearly onto these directions in latent space.  
Because RG evolution mixes semileptonic operators between the electroweak and TeV scales, the corresponding SMEFT-induced distortions trace curved trajectories on the manifold; the latent representation captures these non-linear flows without requiring explicit linearization around a reference scale.
Moreover, since the high-$m_{\ell\ell}$ tail is dominated by contact interactions scaling 
as $\hat{s}/\Lambda^{2}$, the latent model automatically separates interference-driven and 
quadratic contributions, which appear as geometrically distinct curvature directions in the 
embedded manifold.
In this way, the learned latent space provides a compact, physics-aligned coordinate system 
that reflects the inherent properties of the SMEFT, enabling both analytic 
interpretability and computationally efficient inference across the full parameter space.

Taken together, these considerations demonstrate that a foundation model equipped with a physics-aligned latent representation may offer a principled and scalable framework for exploring implications and constraints to SMEFT from Drell--Yan or other hadronic data, complementary to traditional global (EFT) fits.
By learning the manifold on which SMEFT predictions reside, such models encode the full correlation structure of experimental uncertainties, the geometric organization of operator-induced deformations, and the RG-driven evolution linking electroweak and multi-TeV scales. 
This may resolve longstanding limitations of conventional template-based fitting strategies, which struggle to faithfully represent high-dimensional likelihoods, disentangle degenerate directions, or propagate truncation-dependent uncertainties in a consistent manner.
Moreover, the compressed latent formulation enables fast evaluation of theory predictions, 
circumventing the computational bottleneck associated with Monte Carlo generation across 
large Wilson-coefficient grids, a key challenge for HL-LHC precision programs.
Foundation models thus provide a unified and statistically rigorous approach capable of 
integrating Drell--Yan measurements with electroweak precision data, Higgs observables, and 
future collider inputs, paving the way for next-generation analyses that fully exploit 
the information content of available processes.
The demonstrator foundation model presented below represents an initial step toward this larger
potential framework.

\section{Foundation Models}
\label{sec:foundation_models}

There has been a dramatic increase in machine learning-based models in particle physics applied to specific tasks, where a model is constructed and trained on a specific domain dataset relevant for a particular problem. These models generate high-precision predictions for the relevant problem at hand, but their narrow physics reach makes it difficult to reuse them for additional applications. It is also difficult to interpret whether these models have actually learned the physics which they are attempting to predict, or whether they have learned to satisfy some loss metric which is optimized during training in which the two may be in tension with one another. In these cases, it is typically inconsequential whether these models have actually learned physics because they are steered by physicists through domain specific analyses such as a global fit. However, if the aim is autonomous physics discovery --- where a neural agent is potentially steering the analysis and the model is effectively uncovering new correlations in the training corpus --- there is now a necessity to ensure that the models have understood the proper physics. This is the realm of physics-based foundation models.

\begin{figure}
    \centering
    \includegraphics[width=0.9\linewidth]{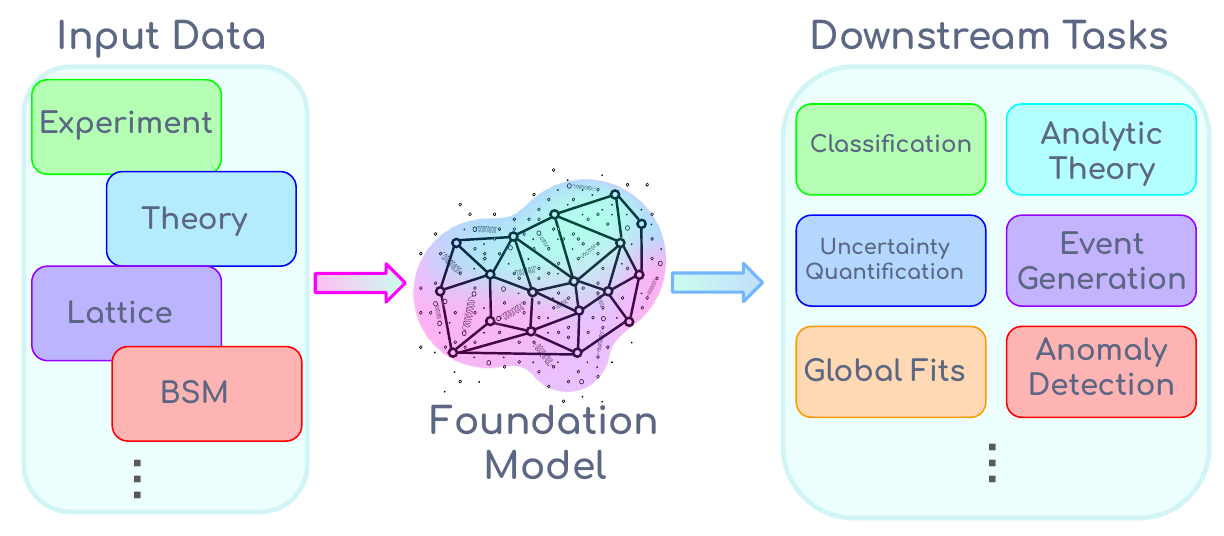}
    \caption{Illustration of a fundamental physics foundation model which ingests large amounts of multi-modal inputs from experiment, theory, lattice, and BSM theories and performs a wide variety of downstream tasks such as classification, uncertainty quantification, global fits, anomaly detection, etc. The foundation model constructs a depiction of the underlying theory through neural connections.}
    \label{fig:foundation_model}
\end{figure}

Foundation models offer an exciting new perspective for autonomous New-Physics discovery in high-energy physics theory. The ultimate goal is to develop a model which is broadly pretrained on a large amount of data, which can then be compressed into an embedding space where the model uncovers fundamental New-Physics correlations between these datasets through training with a variety of typically unsupervised pretraining tasks (see Fig.~\ref{fig:foundation_model} for a diagrammatic explanation). The language of these foundation models will be the same language in which we interrogate our theoretical models in traditional physics, through the lens of collider phenomenology. The foundation model will then be able to learn beyond inference or other tasks and instead be able to understand at a base level the underlying theory in which it is being trained on. From this highly nontrivial embedding space, the model will be able to perform several downstream tasks on which is was not originally trained.

In this section, we give a brief overview of the fundamentals of foundation models and how they can be applied to problems in high-energy theory. We discuss our specific use-case including the data generation process and training corpus, the model specific details such as the pretraining tasks, and the downstream applications of the resulting embedding space.

\subsection{Overview}
\label{sec:overview}

Machine learning has quickly been adopted as a standard tool for collider phenomenology, complementing a wide array of supervised and unsupervised physics tasks such as precision QCD measurements within the SM, collinear hadronic structure, and searching for BSM physics signatures~\cite{Allaire:2023fgp, Gao:2022srd, Kriesten:2025gti, Kriesten:2024ist, Kriesten:2024are, Almaeen:2024guo, Kriesten:2023uoi,Almaeen:2022imx,Grigsby:2020auv}. In many cases, these ML models are defined within a specific physics scope, often investigating a single task in which to outperform standard computational and statistical methods and are trained to minimize an error metric rather than on the explicit physics structure itself. Many challenges have been confronted in recent interrogations of the physics content of such models: the learned latent representation can potentially rely on spurious non-physical correlations for data generation, and these representations are very difficult to interpret in terms of the underlying physics constraints~\cite{Kriesten:2023uoi}. These challenges motivate the construction of large pretrained models with reusable latent embeddings which can support various downstream phenomenological tasks. 

This manuscript will focus on foundation models, which in many ways tackle the above challenges by construction. Within high-energy physics there has been increased attention to foundation models trained on the highly structured collider experimental data spanning event- and jet-level observables and varying detector modalities. A recent review on physics-based foundation models can be found in Ref.~\cite{Hallin:2025ywf} and the references therein. In the rest of this subsection we will detail three main areas of interest of the collider-based foundation model literature: adapting LLMs to experimental workflows, building physics-based foundation models from scratch, and agentic systems for end-to-end autonomous discovery.

\noindent \textbf{Adapting LLMs for collider physics workflows.} General purpose language-based foundation models or large language models (LLMs) have been recently investigated for use in high-energy physics workflows. In such cases, the physics data is translated into text-like formats to be ingested by the pretrained LLM and fine-tuned to perform downstream tasks such as event classification, anomaly detection, and analysis automation \cite{Wildridge:2024yeg, Zhang:2024kws, Vigl:2024lat, Sagar:2025mrd, Saqlain:2025owc,Heneka:2025fpe, Barman:2025wfb}. Related efforts have developed novel tokenization schemes for mathematical reasoning \cite{golkar2024xvalcontinuousnumericaltokenization}. In this case a continuous embedding allows for an improved handling of floating point numbers by LLMs, showing a marked decrease in mathematical hallucinations. The above approaches are particularly attractive in that they preserve the LLMs capabilities (such as code generation, text / image generation, tool use) while still addressing the physics constraints of the experiments.

\noindent \textbf{Building physics-based foundation models from scratch.} Particular attention has been paid to building physics-native foundation models from scratch which are domain-specific to collider phenomenology and experimental data. These models are typically trained using event generation and autoregressive tasks such as next-token or masked-token prediction. Particular examples are anomaly detection through tokenization at the LHC~\cite{Visive:2025flm}; the ``Omni'' series such as OmniLearn \cite{Mikuni:2024qsr,Mikuni:2025tar}, and OmniJet jet foundation models \cite{Birk:2024knn,Bhimji:2025isp} for jet classification and event-level reconstruction; various jet-centric foundation models which emphasize transfer learning across datasets \cite{Leigh:2024ked, Ho:2024qyf, OmanaKuttan:2024mwr, Qu:2022mxj, Harris:2024sra}; predictive embedding schemes like HEP-JEPA \cite{Bardhan:2025icr}; Lorentz symmetry preserving scalable architectures L-GATr \cite{Brehmer:2024yqw}; as well as experimental-based foundation models spanning many detector modalities as in the FM4NPP project \cite{Park:2025ebs}. A complementary direction focuses on building universal latent spaces in which to embed BSM models through their phenomenological similarity \cite{Hallin:2024gmt}.

\noindent \textbf{Agentic systems for autonomous discovery.} A third direction has been developing autonomous agentic systems for New-Physics discovery. Recent investigations have used agents for symbolic regression and learning analytic forms for the underlying physics \cite{Song:2025odk, Morales-Alvarado:2025isx}, as well as using agents for cross domain particle physics discovery \cite{Bakshi:2025fgx, Diefenbacher:2025zzn}.

This manuscript focuses on theoretical `signature-level' foundational embeddings, with an emphasis on semantic interpretability of the latent geometry. SMEFT-based deformations of the neutral-current Drell--Yan process are embedded into a shared representation built from differential distributions with controlled uncertainties. The latent space over SMEFT universes can then be directly exploited for a host of downstream tasks including interpretability clustering, classification, anomaly detection, and nearest-neighbor retrieval.

\subsection{Data-Generation and Training Corpus}
\label{sec:data}

In this section we describe the data-generation process underpinning our foundational embedding space, including event generation and the modeling of effective uncertainties as implemented in the \texttt{PRAXIS} algorithm~\cite{praxis}, the construction of the full training corpus together with an effective tokenization scheme, and the various modeling assumptions that define the boundaries of our foundation model framework for SMEFT deformations of neutral-current Drell--Yan production, $p\,p \rightarrow \mu^{+}\,\mu^{-}$.

\subsubsection*{Data Assumptions and Scope} In this demonstrator foundation model analysis we define the boundaries of the foundational embeddings through the language of collider phenomenology. We limit the discussion to the neutral-current Drell--Yan process, $p\,p \rightarrow \mu^{+}\,\mu^{-}$. We consider proton-proton collisions at a center-of-mass energy of $\sqrt{s}=13$ TeV relevant for current and future LHC and HL-LHC run scenarios. Our input kinematical distributions are the invariant mass, $m_{\ell \ell}$, of the final-state dilepton pair, and the transverse momentum, $p_{T}$, of the muon $\mu^{-}$. These differential distributions are evaluated in $10$ GeV bins over the ranges $m_{\ell \ell} \in [50, 650]$ GeV, and $p_{T} \in [10, 325]$ GeV. We implement acceptance cuts on the rapidity, $\eta$, such that $|\eta| < 2.5$ and the transverse momentum of the muon $p_{T} > 10$ GeV. More information regarding the construction of the datasets and their relevant uncertainties will be given below. The QCD, EW, and SMEFT contributions to the differential cross sections are calculated at leading order (LO). The input PDFs are frozen and taken from the CT18LO PDF set. The QCD parameters such as $\alpha_{s}$ and the heavy quark masses are taken from the corresponding LHAPDF set.\footnote{In order to assess the numerical impact of PDF flavor--scheme choices on our predictions, we compared the Drell--Yan cross section for $p\, p \to \mu^+\mu^-$ using PDF sets with different numbers of active quark flavors (the default 4-flavor scheme versus a 6-flavor scheme). For the largest Wilson--coefficient benchmark considered ($C_{\ell q}^{(1)} = 2$), the corresponding variation in the total cross section is at the level of \textbf{1.5\%}. This spread is significantly smaller than the dominant theoretical systematics in our analysis, such as PDF uncertainties within a fixed scheme and SMEFT truncation effects, ignoring contributions of $\mathcal{O}(1/\Lambda^{4})$. We therefore regard the dependence on the choice of active--flavor number in the PDF evolution as subdominant for the purposes of this study; future SMEFT foundation models can, however, admit the evolution and flavor scheme choices into a wider meta-data register.}
We truncate the SMEFT expansion to $\mathcal{O}(\Lambda^{-2})$, the linear interference contribution from the dimension-6 SMEFT operators with the SM is defined by the SMEFT UFO \smeft (\texttt{SMEFTsim\_U35\_MwScheme\_UFO}) \cite{Brivio:2017btx,Brivio:2020onw}. In our calculations here, we consistently take the New-Physics scale to be $1$ TeV. The SMEFT Wilson coefficients are varied log-uniformly within specified bounds across the board, as described below. Finally, we model effective per-bin uncertainties using a Gaussian-kernel prescription, which will be detailed in the following subsection.

\subsubsection*{Event Generation} The \texttt{PRAXIS} algorithm was developed to systematically explore the SMEFT Wilson coefficient parameter space for large-scale phenomenological studies such as a foundation model. In its current implementation, it imposes a binary decision rule that identifies which SMEFT Wilson coefficients contribute to a specific collider process at a specified order in the SMEFT expansion. In practice, \texttt{PRAXIS} steers an implementation of \MG to toggle individual Wilson coefficients and checks whether the resulting cross section is non-zero. This algorithm samples the Wilson-coefficient space and, at $\mathcal{O}(\Lambda^{-2})$, returns the subset of operators whose interference with the SM amplitude yields a non-vanishing contribution to the cross section. It should be noted that this subset does not necessarily coincide with the set of operators which enter the particular underlying amplitudes. At $\mathcal{O}(\Lambda^{-2})$, the interference contribution to the total cross section is 
\begin{eqnarray}
    \sigma_\mathit{int} &=& 2\,\Re e(\mathcal{A}_{SM}^{*}\, \mathcal{A}_{d6}) = 2\,[\Re e (\mathcal{A}_{SM})\,\Re e(\mathcal{A}_{d6}) + \Im m (\mathcal{A}_{SM})\, \Im m (\mathcal{A}_{d6})]
\end{eqnarray}
For certain choices of operators and phase-space configurations this real part can vanish, such that $\sigma_\mathit{int} = 0$ even though $\mathcal{A}_{d6} \neq 0$. Since these contributions do not modify the phenomenologically accessible differential cross sections used as inputs to our foundation model, we do not consider them in this study. For neutral-current Drell--Yan, the application of \texttt{PRAXIS} at $\mathcal{O}(\Lambda^{-2})$ thus defines the subset of Wilson coefficients over which we subsequently sample to construct individual SMEFT universes, as described below.

For neutral-current Drell--Yan, the application of the \texttt{PRAXIS} algorithm at $\mathcal{O}(\Lambda^{-2})$ defines a subset of Wilson coefficients with non-zero contribution to the cross section over which we sample to construct individual SMEFT universes. For each coefficient $C_{i}(\mu)$ in this defined subset we first set the magnitude by drawing from a log-uniform distribution,
\begin{eqnarray}
    \ln |C_{i}(\mu)| \sim \mathcal{U}(\ln{a}, \ln{b})\, , \qquad a = 10^{-4}, \; b=2\, ,
\end{eqnarray}
and then fix the associated sign randomly through draws of a binomial distribution. This ensures that the Wilson coefficients have a broad support and are symmetric in sign over the interval $[10^{-4}, 2]$. Finally, we introduce sparsity in a controlled manner for a physics-based, non-trivial sampling of the full parameter space. For each Wilson coefficient we draw a random variable,
\begin{eqnarray}
    p \sim \mathcal{U}(0,0.4)\, ,
\end{eqnarray}
which defines the probability to set each individual coefficient to zero. The purpose of this random sampling, as apposed to grid-based scans, is to better mimic the patterns which come from theoretical matching of UV-complete models onto the SMEFT parameter space. Sample grid-based searches that turn on and off Wilson coefficients one at a time generate scenarios that are difficult to realize in concrete UV completions. By contrast, our random sampling naturally admits matching configurations in which each SMEFT universe corresponds to some collection of heavy New-Physics scenarios.

We construct 100 of these SMEFT universes, each of which defines a specific realization of the neutral-current Drell--Yan process and manifests as deviations from the SM kinematical distributions. We use the \MG event generator to simulate the hard-scattering process with the relevant SMEFT Wilson coefficients tuned to their corresponding values. The result is a set of input training data consisting of high-resolution singly-differential distributions in $m_{\ell \ell}$ and $p_{T}$ generated by iterating over kinematic bins in $10$ GeV increments and collecting the differential cross-section value for each bin. We use the CT18LO PDFs as described above and set the invariant mass limits from $m_{\ell \ell} \in [50, 650]$ GeV and the transverse momentum $p_{T} \in [10, 325]$ GeV resulting in $N_{m_{\ell \ell}}\! +\! N_{p_{T}} = 92$ input dimensions per SMEFT universe. Thus, at this stage, our dataset consists of $100$ theory-level kinematical distributions in a $92$-dimensional input space corresponding to different realizations of varied SMEFT Wilson coefficients. The next step in the data-generation pipeline is to translate these theory curves into a training corpus of Monte Carlo replicas with physics-motivated, simulated uncertainties.

\subsubsection*{Modeling Uncertainties} To generate Monte Carlo replicas from the high-resolution theory curves, we first need to evaluate the per-bin statistical uncertainty for each differential distribution used to train the foundation model. Conceptually, we model these uncertainties as arising from a mock experiment with an effective integrated luminosity, $L_\mathit{eff}$. We set this scale by fixing a target number of expected events and performing a single baseline \MG run with $10^4$ unweighted events for the neutral-current Drell--Yan process, binning these into $p_{T}$ and $m_{\ell \ell}$ differential distributions. This results in an order $\mathcal{O}(10)$ events in the high-mass tail bins. The output provides an estimate of the total cross section, $\sigma_\mathit{tot}$, along with the Monte Carlo statistical uncertainty, $\Delta \sigma_\mathit{tot}$. This total integrated uncertainty we interpret  as the statistical fluctuations corresponding to an effective total integrated luminosity, $L_\mathit{eff}$, and use it to calibrate a per-bin variance which we detail below.

First, we assume that the number of events per-bin follows a Poisson distribution, as is generally consistent for the statistics of empirical measurements like differential cross sections inferred from counts. To first approximation, we treat each scattering event as an independent Bernoulli trial, where the number of rare events, $k$, in a given bin is given by the Poisson limit of the Bernoulli/Binomial distribution,
\begin{eqnarray}
    \text{Pr}[N=k] = \text{Pois}(k; \mu) = e^{-\mu}\,\frac{\mu^{k}}{k!}\, ,
\end{eqnarray}
where $N$ is the Poisson distributed random variable denoting the number of events and $\mu$ is the mean. For our demonstration case of neutral-current Drell--Yan, the differential cross section is the largest in bins near the $Z$-pole (neglecting contributions at $0$ from the photon propagator), and therefore contains the most events and smallest relative uncertainties in the corresponding bins; meanwhile, bins in the tails of distributions contain far fewer events and are hence characterized by much larger relative uncertainties.
We emphasize that the Poissonian model used here is not intended as a representation of a specific detector-level measurement but serves as a systematically
improvable proxy for realistic uncertainties with explicit luminosity dependence.

With this Poisson distribution assumption, the expected number of events in bin $i$, $\mu_{i}$, grows linearly with the value of the per-bin cross section weighted by the total integrated luminosity,
\begin{eqnarray}
    \mu_{i} = L\, \sigma_{i}\ ;
\end{eqnarray}
here, we ignore detector effects such as bin-dependent efficiencies. Denoting the differential cross section in bin $i$ by $y_{i} = (d\sigma / dX)_{i}$ and the bin width as $\Delta X_{i}$, we have $\sigma_{i} = y_{i}\,\Delta X_{i}$ such that
\begin{eqnarray}
    \mu_i = L\,y_i\,\Delta X_i \, , \qquad
    N_i \sim \mathrm{Pois}(\mu_i)\, .
\end{eqnarray}
By construction, the Poisson mean and variance are both $\mathbb{E}[N_{i}] = \text{Var}[N_{i}] = \mu_{i}$. The differential cross section in each bin can be written as a statistical estimator ({\it i.e.}, a random variable which is a function of another random variable --- denoted here by a ``hat,'' $\hat{a}$):
\begin{eqnarray}
    \hat y_i = \frac{N_i}{L \,\Delta X_i}\, .
\end{eqnarray}
Using the definition of the variance, $\text{Var}[aX] = a^2 \text{Var}[X]$, we obtain
\begin{eqnarray}
    \text{Var}[\hat{y}_{i}] 
    = \text{Var}\left[ \frac{N_{i}}{L\Delta X_{i}}\right] 
    = \frac{\mu_{i}}{L^{2}\Delta X_{i}^{2}} 
    = \frac{y_{i} L \Delta X_{i} }{L^{2}\Delta X_{i}^{2}} 
    = \frac{y_{i} }{L\Delta X_{i}}\, .
\end{eqnarray}
The variance of the cross section estimator, $\hat{\sigma}_\mathit{tot}$, can then be written as
\begin{equation}
    \text{Var}[\hat\sigma_{tot}]
    = \text{Var}\left[\frac{\sum_{i}N_{i}}{L} \right]
    = \frac{1}{L^2} \sum_i \text{Var}[N_i]
    = \frac{1}{L^2} \sum_i \mu_i
    = \frac{1}{L^2} \sum_i L \sigma_i
    = \frac{\sigma_\mathit{tot}}{L}\, .
\end{equation}
For this study, we make an assumption that this variance is represented by the Monte Carlo estimate of the total integrated cross section uncertainty from the \MG event generation, $\Delta \sigma_{tot}^{2} = \text{Var}[\hat\sigma_\mathit{tot}]$. From this total uncertainty assumption we can then define an effective luminosity,
\begin{eqnarray}
    L_{\mathit{eff}}
    = \frac{\sigma_\mathit{tot}}{\Delta \sigma_\mathit{tot}^{2}}\, ,
\end{eqnarray}
which we treat as a global calibration parameter, rather than a bin-by-bin quantity.
Using this effective luminosity, $L_\mathit{eff}$, the per-bin uncertainty of the differential cross section is then: 
\begin{eqnarray}
    \Delta y_{i} = \sqrt{\text{Var}(y_{i})} = \sqrt{\frac{y_{i}}{L_\mathit{eff} \Delta X_{i}}}\ .
\end{eqnarray}
The fractional per-bin uncertainty of the differential cross section,
\begin{eqnarray}
    \frac{\Delta y_{i}}{y_{i}} = \sqrt{\frac{1}{L_\mathit{eff}\, y_{i}\,\Delta X_{i}}} \, ,
\end{eqnarray}
contains an important property of the uncertainty modeling, that the per-bin uncertainty grows as the differential cross section $y_{i}$ decreases, which is what one would expect for a realistic statistical model of experimental measurements. An additional advantage of this construction is that this uncertainty can be effectively rescaled to any target luminosity, $L_\mathit{desired}$. Since the statistical uncertainties scale as $1/\sqrt{L}$, the rescaling factor for the uncertainty can be defined through the effective luminosity as:
\begin{eqnarray}
    m = \sqrt{\frac{L_\mathit{eff}}{L_\mathit{desired}}}\, ,
\end{eqnarray}
relating the two uncertainties at the two different luminosities via $\Delta \sigma_\mathit{tot}' = m\Delta \sigma_\mathit{tot}$, and analogously for the per-bin uncertainties, $\Delta y_{i}$.

With these physics-based per-bin uncertainties for each differential cross section, it is then necessary to model the full covariance matrix across the kinematic bins in order to sample smooth Monte Carlo replicas for training. By relaxing the assumption that the bins are statistically independent, we impose correlated fluctuations through a Gaussian kernel so that the resulting replicas resemble smooth physics distributions rather than non-physical ``jagged'' ones which would otherwise result from randomly sampling diagonal variances. This is important for ensuring that the foundation model learns the underlying smooth structure of the input data distributions. We model the covariance matrix using a Gaussian (radial basis function, RBF) kernel. For a given observable, $X$, with bin centers, $x$ and $x'$, this can be written as 
\begin{eqnarray}
    K(x,x') = \exp{(-\gamma || x - x'||^{2})}\, , \qquad \text{where} \,\, \gamma = \frac{1}{2\ell_{X}^{2}}\, ,
\end{eqnarray} and $\ell_{X}$ sets the correlation length across bins in $X$. In this demonstration, we set a common value of $\ell_{X} = 35$ GeV for both the $p_{T}$ and $m_{\ell \ell}$ differential distributions. Given the per-bin variances, $\text{Var}[y_{i}]$, for $y_{i} = (d \sigma / d X)_{i}$, we construct the initial statistical covariance matrix as 
\begin{eqnarray}
    \Sigma_{o} = D \, K(x,x')\, D\, , \qquad \text{where} \,\,D = \text{Diag}\left(\sqrt{\text{Var}[d\sigma_{i} / dX]}\right)\, ,
\end{eqnarray}
so that the off-diagonal covariances are $(\Sigma_{0})_{ij} = \sqrt{\text{Var}[y_i]\,\text{Var}[y_j]}\,K_{ij}$ and the diagonal terms match the per-bin statistical framework we set up above $(\Sigma_{0})_{ii} = \text{Var}[y_i]$.

The variance on the integrated cross section associated with a covariance matrix $\Sigma$ on the binned differential cross section is
\begin{eqnarray}
    \text{Var}[\hat{\sigma}_{tot}] = \text{Var}\left[\sum_{i} y_{i} \Delta X_{i} \right] = \Delta X^{\top} \Sigma \, \Delta X\, ,
\end{eqnarray}
where $\Delta X$ denotes the column vector of bin widths, $\Delta X_{i}$ (which in this case is a constant vector). The off-diagonal covariances encoded in $K$ modify the variance of the integrated cross section from its diagonal-only Poisson value as calculated above. To ensure consistency with the Monte Carlo estimate, $\Delta \sigma_\mathit{tot}^{2}$, from \MG, we must rescale the covariance matrix by a global factor,
\begin{eqnarray}
    \Sigma = \alpha \Sigma_{o}\, , \qquad \text{where}\,\, \alpha = \frac{\Delta \sigma_\mathit{tot}^{2}}{\Delta X^{\top}\Sigma_{o} \Delta X}\, .
\end{eqnarray}

Finally, for each SMEFT universe we draw smooth replicas of the differential distributions by sampling from a multivariate normal distribution with mean given by the theory calculation and the covariance matrix $\Sigma$ defined above,
\begin{eqnarray}
    R_{i} = \mathcal{N}(\mathbf{y}, \Sigma)\, ,
\end{eqnarray}
where $\mathbf{y}$ is the column vector of binned differential cross section values $y_{i} = (d \sigma / dX)_{i}$.

\begin{figure}
    \centering
    \includegraphics[width=0.495\linewidth]{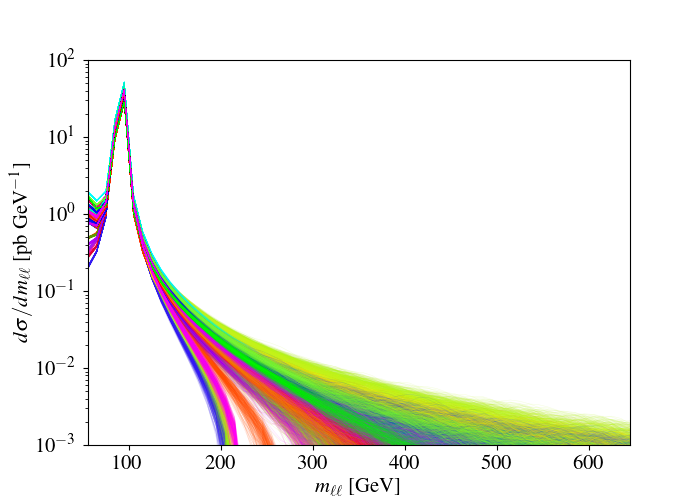}
    \includegraphics[width=0.495\linewidth]{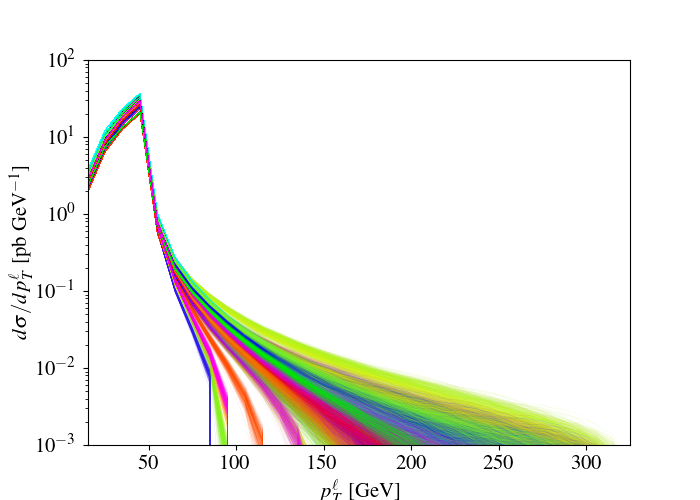}
    \caption{Examples of the input differential distributions for the neutral-current Drell--Yan process with $100$ ``SMEFT'' universes resulting in deviations from the SM predictions in ({\it left}) the invariant mass distribution $m_{\ell \ell}$, and ({\it right}) the transverse momentum distribution of the muon $p_{T}$.}
    \label{fig:placeholder4}
\end{figure}

\subsubsection*{Tokenization} Transformer-based foundation models typically ingest training data through a sequence of discrete tokens, requiring an explicit tokenization scheme mapping raw data inputs into a form which would be suitable for the encoder. Typical tokens are effectively vectors of information which contain specific meta-data regarding the types of inputs which would be necessary for the pretraining tasks. In this case such tokens would possibly contain, in addition to the kinematical dependence of the differential distributions, information regarding the perturbative order of the hard-scatter calculation, the values for the strong coupling or other perturbative inputs like the heavy-quark masses, the SMEFT truncation order, the separation of higher dimension operators, or even the specific PDFs used which inform parton-level luminosities. The tokenization scheme becomes a very important step when building a foundation model as it dictates what types of pretraining tasks one can perform, and therefore the maximum correlations the foundation model can ultimately create. 

In this demonstrator study, each SMEFT universe is represented directly as a fixed length vector of concatenated binned differential cross sections in $m_{\ell \ell}$ and $p_{T}$ for a single collider process with all of the assumptions we detailed in Sec.~\ref{sec:data}. Therefore, there is no need in this first study for the foundation model to semantically separate the latent space on the training meta-data. Thus, in this scenario we employ a feed-forward neural architecture as our encoder model rather than a transformer-based encoder; as such, the network can already ingest these vectors directly without a separate tokenization or embedding scheme.
We note that, in extensions of this calculation, the generalization to a transformer-based architecture can be carried out while retaining the essential
features of the present foundation model.

\subsection{Model Architecture and Contrastive Pretraining}
\label{sec:arch}

In this section we describe the encoder architecture used to embed the differential distributions from Sec.~\ref{sec:data}, the key hyperparameters of our model and the rationale behind their choice, and the supervised contrastive pretraining task used during training to separate the various SMEFT universes in the latent space.

\subsubsection*{Model Architecture and Training} The encoder model we employ to construct our foundation embeddings consists of standard feed-forward fully connected blocks with the following pattern: 
\[
    \text{input} \;\rightarrow\; \mathrm{Dense}(100,\,\text{L2 kernel reg.} = 0.1)
    \;\rightarrow\; \text{BatchNorm}
    \;\rightarrow\; \text{ReLU}
    \;\rightarrow\; \text{Dropout}(p = 0.35)\, .
\]
Here, $\mathrm{Dense}(\cdot)$ is a fully connected Dense layer with L2 kernel regularization that penalizes the sum of the squares of the weight parameters to prevent overfitting; \text{ReLU} is the rectified linear unit activation function; the BatchNormalization layer normalizes the activation in each mini-batch to have mean zero and a standard deviation of one; and $\text{Dropout}(\cdot)$ is a regularization layer which independently sets activations to zero with probability equal to $p$. The encoder consists of three such feed-forward blocks followed by an output Dense layer that projects to the latent space. In this case, we take the dimensionality of the encoded representation to be $2$. The encoder model is trained using the Adam optimizer with a learning rate of $10^{-6}$, and early stopping to prevent overfitting by monitoring a hold-out validation dataset.

For the purposes of this demonstration, we deliberately adopt a minimal encoder architecture in order to test the impact of the SMEFT-induced deviations. Since most of our downstream tasks involve visualization and interpretation of the latent space, we fix the dimensionality of the foundation embedding to $2$ so that this can be performed directly without an additional projection. In preliminary exploratory tests we trained models with higher-dimensional latent spaces and subsequently projected the latent embedding using principal component analysis (PCA); the resulting geometry of the SMEFT universes were qualitatively unchanged. This indicates that, for the present dataset, a two-dimensional latent captures dominant variances in the training data inputs. Similarly, we found that an encoder depth of three fully connected feed-forward blocks gave the best trade-off between expressivity and generalization; with deeper variations of the encoder model tending to overfit immediately. This also explains the high levels of regularization in the model with the inclusion of Batch Normalization layers, Dropout, and the L2 kernel regularization. A more systematic exploration of possible larger architectures and higher-dimensional embeddings is left to future work.

Given the nature of the contrastive pretraining task which we will detail below, we must organize the dataset into positive and negative pairs of examples for the contrastive loss. Starting from the Monte Carlo replicas, this is done by treating each SMEFT universe as a distinct class: there are $100$ of these universes and for each of them we generate $10^4$ replicas, resulting in a total of $10^{6}$ total training examples. Each replica is associated with an integer class label $c \in \{0,\dots,99\}$ indicating which SMEFT universe ({\it i.e.}, which set of Wilson coefficients) it was generated from in the \texttt{PRAXIS} pipeline. We then shuffle the dataset so that the ordering of the labels is completely randomized while the association between each label and its binned kinematic distribution $(m_{\ell \ell},p_{T})$ is preserved. 

From this labeled pool of randomized data we construct a set of contrastive training pairs. We first specify a total number of pairs to be used and the desired fraction of positive versus negative pairs. For our example, a positive pair consists of two replicas which share the same universe label, $c_{i} = c_{j}$, and therefore correspond to the same underlying SMEFT deformation of the neutral-current Drell--Yan process. The differences between these positive pairs are in their representation which in this case is represented by the statistical fluctuations which are modeled based on the formulation as described in the previous section. A negative pair consists of two replicas with different labels, $c_{i} \neq c_{j}$, and therefore come from different points in the underlying SMEFT parameter space. This will typically lead to distinct patterns in the types of shape deviations in the physics distributions.  We then sample the pairs until the target number of positive and negative pairs are reached, forming a stratified training set of pairs over the initial labels. Each pair is then assigned a new target $y$, where $y=1$ corresponds to positive (same-universe) pairs and $y=0$ corresponds to negative (different-universe) pairs. 

In this way, the contrastive pretraining tasks is aligned with the physics task of distinguishing between SMEFT universes through similarity; the encoder is encouraged to squeeze replicas generated from the same (or similar) SMEFT universe to nearby points in the latent space, while pushing apart the replicas generated from different SMEFT universes corresponding to distinctive theories.

\subsubsection*{Pretraining Tasks} 

We employ a supervised contrastive pretraining task in which the similarity between two training example pairs is defined through the corresponding label of the underlying SMEFT universe. For each pair, $(x,x')$, of Monte Carlo replicas, we assign a binary label $y \in \{0, 1 \}$, where $y=1$ corresponds to positive pairs drawn from the same SMEFT universe and subsequently same Wilson coefficient configuration, and $y=0$ corresponds to negative pairs drawn from different SMEFT universes. The encoder model, which we denote as $E_{\theta}$, is trained with a pairwise contrastive loss defined through the loss function
\begin{eqnarray}
    \mathcal{L}_{\theta}(x,x',m) = y D_{\theta}(x,x')^{2} + (1-y)\max{(0, m-D_{\theta}(x,x'))^{2}}\, ,
\end{eqnarray}
where $D_{\theta}(x,x')$ is the Euclidean distance in the latent embeddings of $x$ and $x'$, and $m$ is a hyper-parameter which sets the separation scale for dissimilar pairs. Denoting $z = E_{\theta}(x)$ and $z' = E_{\theta}(x')$ for the encoded representations of the Monte Carlo replicas in $d$ dimensions, we can define the Euclidean distance in this latent space as,
\begin{eqnarray}
    D_{\theta}(x,x') = \sqrt{\sum_{k=1}^{d}\left( z_{k} - z'_{k}\right)^{2}}\, .
\end{eqnarray}
With this definition, positive pairs ($y=1$) are encouraged to have small distances since $D_{\theta}(x,x')$ is minimized, while negative pairs ($y=0$) are encouraged to separate. This supervised contrastive loss plays a similar role to the contrastive loss used in unsupervised representation learning in which a cosine similarity loss is used to separate the pairs. However, in our case, since the notion of similarity is defined through the SMEFT model label, we can semantically relate the distance in the embedded latent representation to the the SMEFT-based deformations in the differential distributions of the neutral-current Drell--Yan process. 

Recent trends in foundation model research indicates that multi-task pretraining can substantially improve model generalizability. By exposing the model to a diverse set of pretraining tasks, the latent space is encouraged to organize itself into a semantically rich topology complementing each task it has seen, and therefore enhancing performance of the foundation model on downstream tasks which were not explicitly seen during training. In this demonstrator study, we choose a single supervised contrastive pretraining task tailored towards our downstream physics goals of parsing and interpreting the geometry of the latent distribution. However, for a future study it is well motivated to pursue a multi-objective pretraining corpus encompassing many different physics-based objectives. Examples of these types of objectives include: ({\it i}) classifying the perturbative order of contributions (LO {\it vs.}~NLO in SMEFT, QCD, and EW sectors), ({\it ii}) disentangling SM hadronic uncertainties such as those arising from the PDFs from SMEFT deformations in the differential distributions, ({\it iii}) identifying the SMEFT truncation order [{\it i.e.}, distinguishing $\mathcal{O}(\Lambda^{-2})$ {\it vs.}~$\mathcal{O}(\Lambda^{-4})$], and ({\it iv}) generating new, unseen physics distributions from the latent embedding. Training on a suite of such pretraining tasks will encourage the model to learn a richly structured and highly correlated embedding which reflects multiple facets of the underlying SMEFT. 

\section{Results}
\label{sec:results}

In this section we will use the foundation model embedding space to perform three downstream tasks:  classification with uncertainty quantification, anomaly detection quantified with information theoretic metrics, and nearest-neighbor retrieval in the presence of SM uncertainties. We present these results with a suite of visualization techniques that enable us to interpret the foundational embeddings and allow us to extract potential New-Physics content from their learned correlations.

\subsection{Interpreting the foundational embedding}

The embedding is the latent backbone of the foundation model that encodes the relevant physics content along with learned correlations. It is constructed by nonlinear projections of the differential distributions described in Section~\ref{sec:data} into a reduced dimensional space with a contrastive task creating semantic separations between different SMEFT universes. Interrogating the latent geometry furnishes some interpretation of how the embedding model organizes the learned physics correlations. We can demonstrate two interpretability exercises on the latent representation: ({\it i}) we will investigate the organization of the kinematic information among the SMEFT universes by overlaying specific kinematic bins of interest, ({\it ii}) we will unpack the embedded distribution and look at clusters in the embedding and investigate similarities and differences.

\begin{figure}
    \centering
    \includegraphics[width=0.32\linewidth]{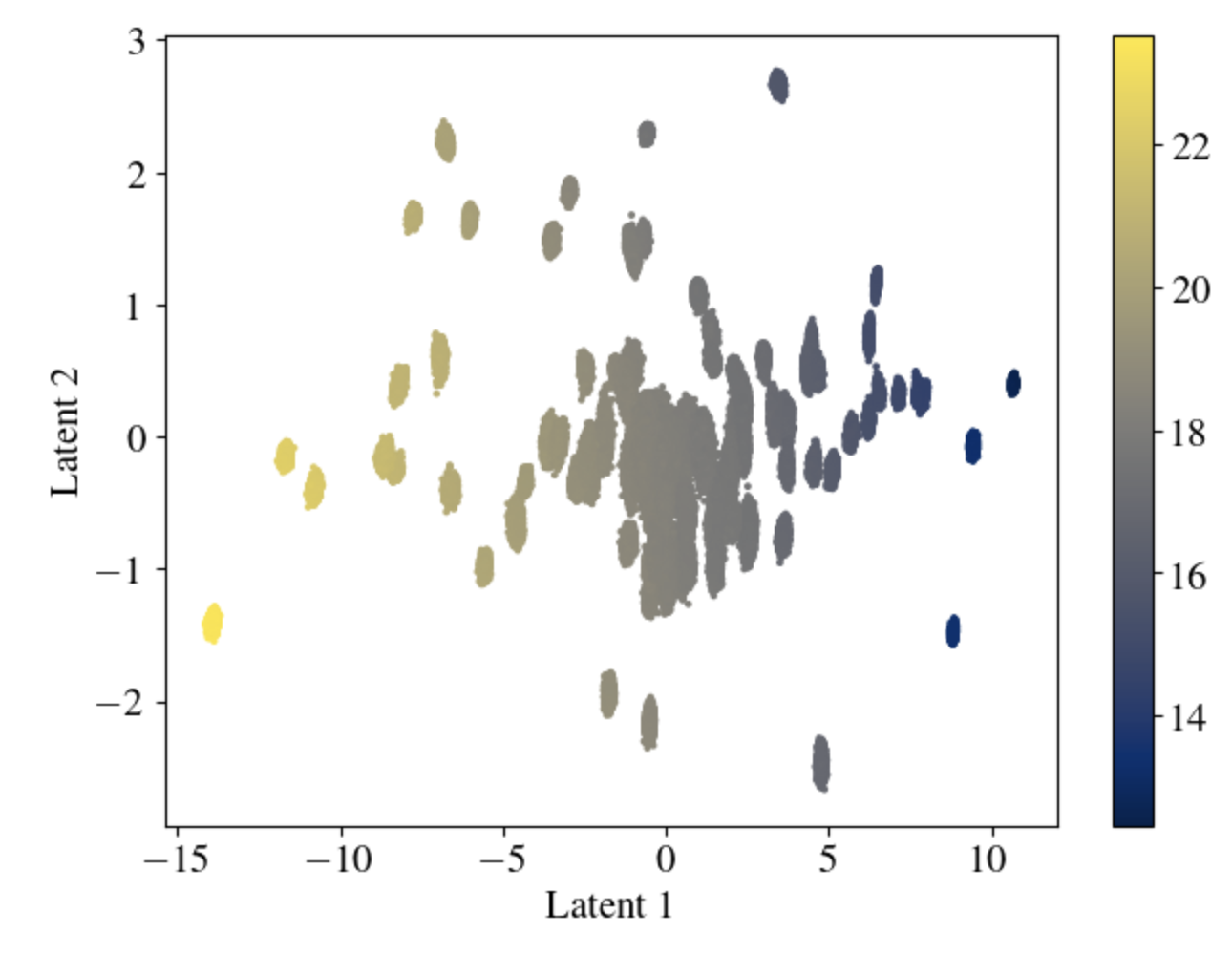}
    \includegraphics[width=0.325\linewidth]{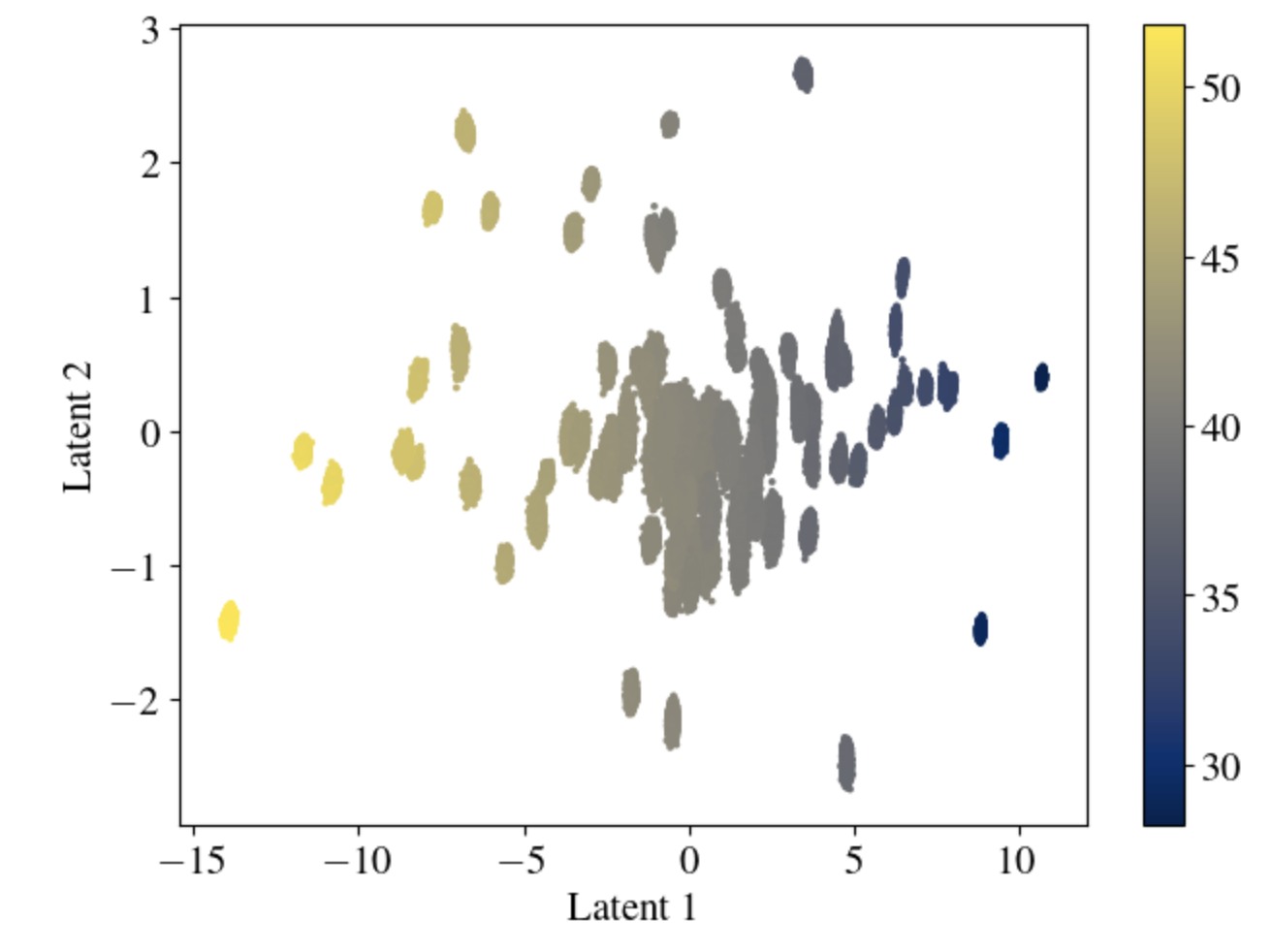}
    \includegraphics[width=0.34\linewidth]{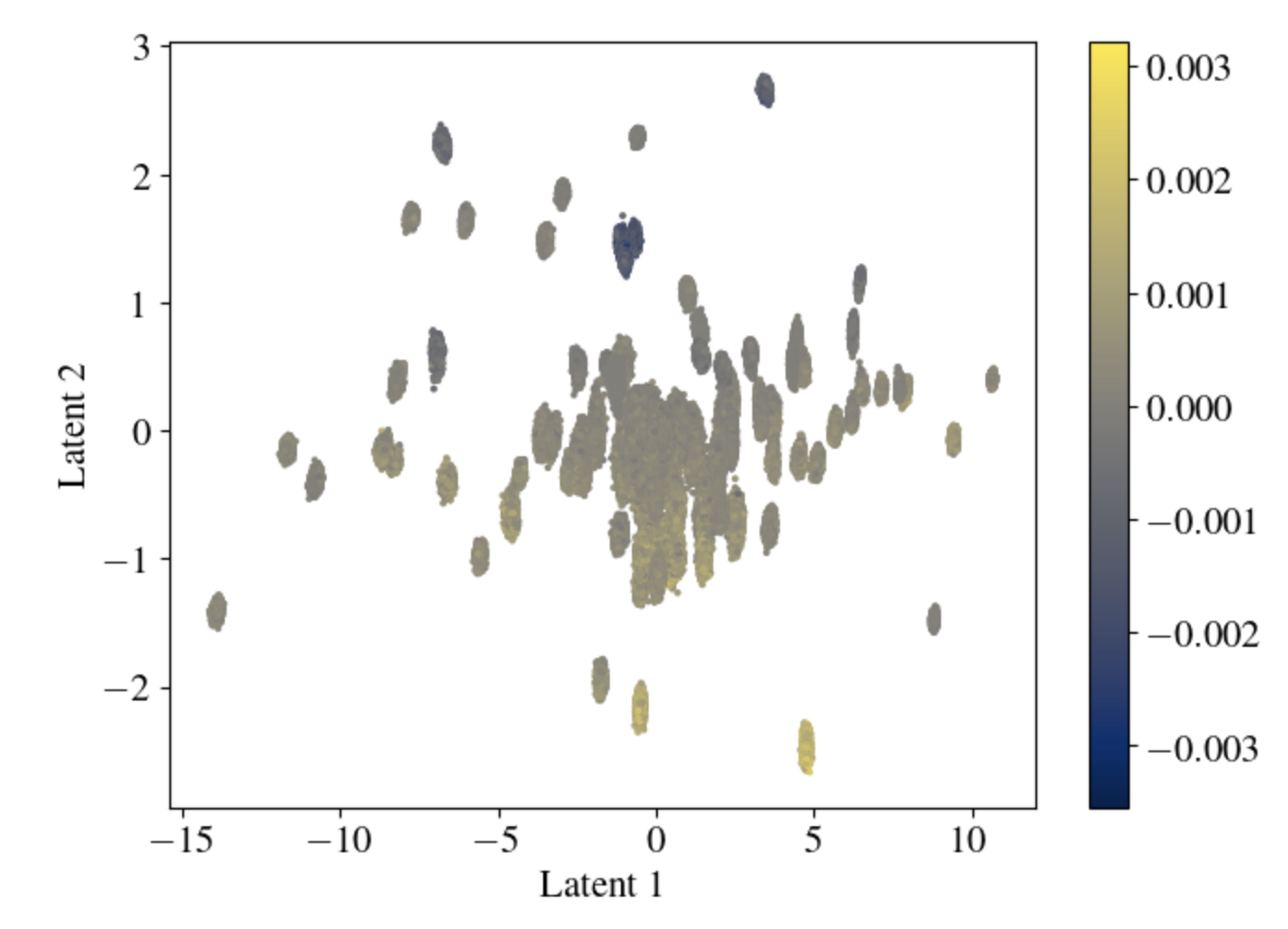}
    \caption{Foundational embeddings of 100 SMEFT universes into a 2D latent space where the color scale quantifies the associated values of three different single-differential cross sections at specific kinematical points. ({\it Left}) scans the value of the transverse momentum spectrum, $d\sigma/dp_T$, in the kinematic bin just below the $Z$-pole; ({\it middle}) scans over the value of the invariant mass distribution of the final dilepton state at the $Z$-pole; and ({\it right}) corresponds to a high-mass bin, also in the invariant mass distribution. From these visualizations one can see how the foundation model organizes information in the latent space.}
    \label{fig:foundation_embedding}
\end{figure}

In Fig.~\ref{fig:foundation_embedding} we show three examples of the embedding space, overlaying singly-differential cross-section values within specific kinematic bins on each embedding. One can see by the gradations in color that these kinematic bins are organized in different manners in the embedding, but in general, showing a degree of contiguity in the organization based on each of the plotted cross-section values. We choose a kinematic bin just below the $Z$-pole in the transverse momentum differential distribution, a kinematic bin on the $Z$-pole for the invariant mass differential distribution, and a high-mass bin in the invariant mass differential distribution. From the resulting 2D distributions, one can see that the foundation model embedding has clear sensitivity around the $Z$-pole in both $p_{T}$ and $m_{\ell \ell}$. The embedding has also learned a general correlation between peaks in the invariant mass distribution and corresponding peaks in the $p_{T}$ spectra, since the model has organized this information in the same way along the Latent 1 axis. This is an example of such learned correlations that the model is not necessarily trained on. One can also see that the organization of the information in the high-mass tail is more subtle, with a much flatter distribution. There are also definitive peaks in this distribution marking more ``anomalous'' contributions in the training data such as regions where the differential distribution could go negative. These solutions seem to cluster themselves together. The information in this case seems to be organized more along the Latent 2 axis.

There also appears to be a very noticeable region where many solutions sit in the center along coordinates $(0,0)$ in the latent dimensions. This region corresponds to much smaller deformations away from the SM. This is because it is much rarer to generate SMEFT deformations that drastically lie away from the SM without large Wilson coefficients. This also requires there to be little destructive interference between Wilson coefficient configurations, which by design of the data generation, seems to be a rare occurrence. This is interesting in itself because random sampling of the Wilson coefficient space suggests that often one returns close to the SM. We leave investigations for a physics-informed sampling of the Wilson-coefficient space to a future work.

\begin{figure}
    \centering
    \includegraphics[width=\linewidth]{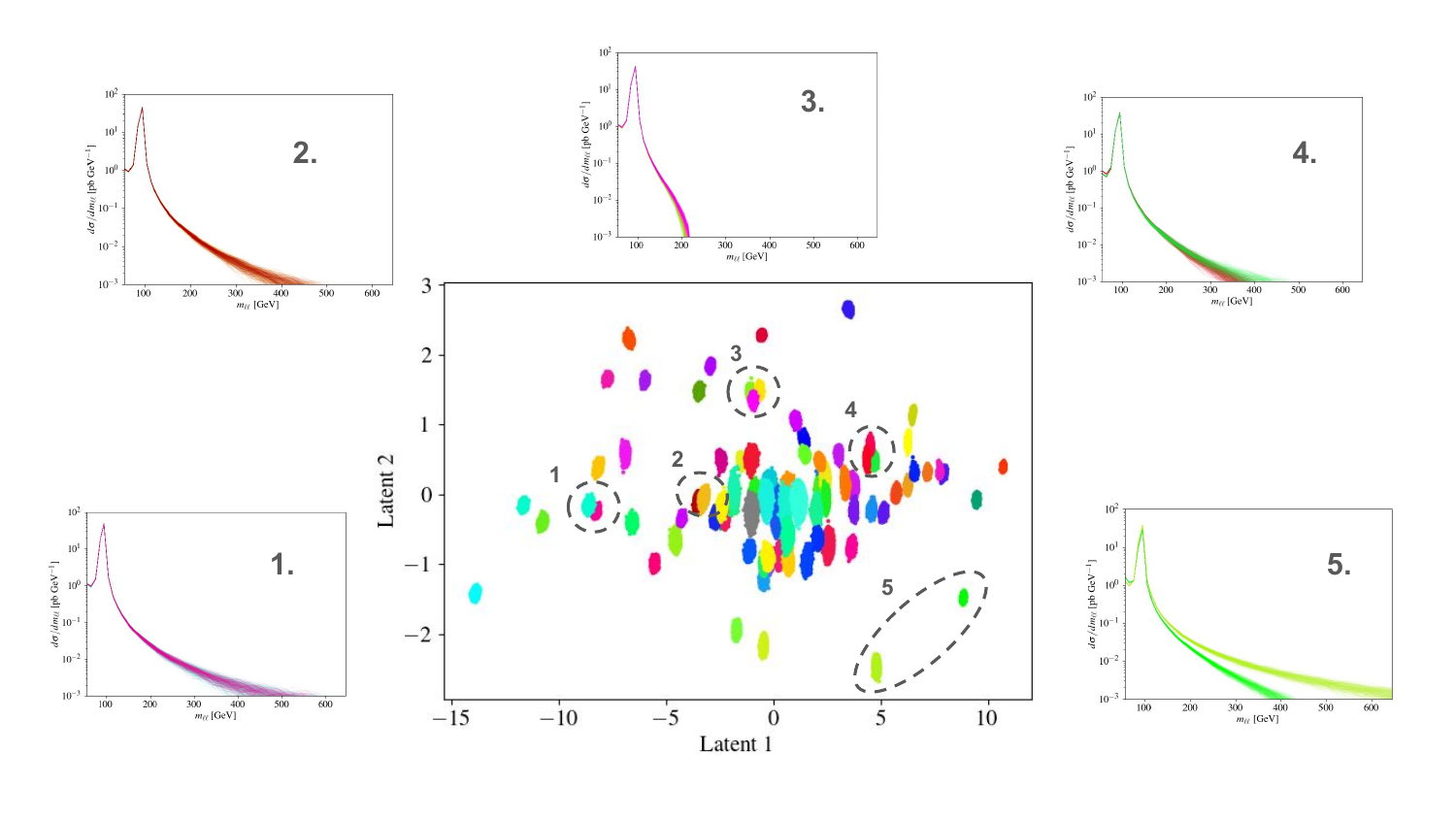}
    \caption{Representative examples of groups within the foundational embedding space demonstrating semantic separation of different behaviors of the invariant mass distribution represented by Euclidean distances.}
    \label{fig:embedding_clustering}
\end{figure}

In Fig.~\ref{fig:embedding_clustering} we interrogate the geometry of the embedding where traversals across the latent space correspond to shape-level changes in the kinematical distributions. This allows us to understand the underlying features that the model is learning and how it organizes that information through clusters of SMEFT universes. We take five samples of the outlier clusters (away from the large SM cluster at the center) to understand how the differential invariant mass distribution morphs as a function of the latent coordinates. In scenarios 1, 2, 3, and 4, we see clusters in which the high-mass tails of the invariant mass distribution change dramatically as one sweeps across Latent 1. We also note that the embedded distributions in these clusters closely resemble one another, hence the overlap in the latent space. In scenario 5, we see an example of semantic separation between two different SMEFT universes showing how the model separates dissimilar SMEFT deformations. Using this latent embedding, we can now perform downstream tasks, several of which we briefly
demonstrate.

\subsection{Downstream tasks}
\label{sec:downstream}

Our downstream tasks center on the identification or interpretation of New Physics as encoded within the foundational embeddings. For this initial study, we focus on classification with uncertainty quantification, anomaly detection, and nearest-neighbor retrieval.

\subsubsection*{Classification with UQ}

We use the foundation model latent embedding to discriminate between SMEFT universes through a trained classification head with uncertainty quantification. Each SMEFT replica is mapped to a latent coordinate where a lightweight classification head outputs Dirichlet parameters for each of the ($C=100$) classes. This method makes use of recent advances in uncertainty estimation using prior network and evidential deep learning techniques \cite{malinin2018predictiveuncertaintyestimationprior, ulmer2023priorposteriornetworkssurvey} where class probabilities are mapped to a latent vector on a simplex, 
\begin{eqnarray}
    \mathbf{\mu} \in \Delta^{C-1} = [\mu_{1},\mu_{2},\dots,\mu_{C}]\,,
\end{eqnarray}
and the predictive distribution is constructed by marginalizing over $\mu$. An ecology of uncertainties can be constructed using the prior network by marginalizing over various parameters: the total predictive uncertainty, separations into epistemic and aleatoric sources, and further segmenting the epistemic into knowledge and distributional uncertainties. Prior networks for classification, in particular the Dirichlet prior network (DPN), have been first used for HEP BSM theory discrimination in neutrino-scattering experiments~\cite{Kriesten:2024ist} as well as in jet-classification studies~\cite{Khot:2025kqg}. Similarly, ongoing work connects the concept of epistemic uncertainties to open questions in PDF-related hadron structure such as the PDF parameterization uncertainty~\cite{Kotz:2025lio}.

Standard statistical methods tell us that the predictive total classification uncertainty can be written as a marginalization over model parameters, $\theta$, and class probabilities, $\mu$, as:

\begin{eqnarray}
    p(y | \mathbf{x}^{*},\mathcal{D}) &=& \int \int p(y|\mu)\, p(\mu | \mathbf{x}^{*},\theta)\, p(\theta | \mathcal{D}) d\theta d\mu\ ,\ \
\end{eqnarray}
where in this case $\mathbf{x}^{*}$ represents the unseen test examples from which one makes a prediction of the class labels $y$ given a total initial dataset $\mathcal{D}$. We can take the approximation $p(\mu | \mathbf{x}^{*}, \mathcal{D}) \approx p(\mu | \mathbf{x}^{*}, \hat{\theta})$ and now define a neural network which maps a Dirichlet prior distribution over class probabilities by predicting concentration parameters, $\alpha$,

\begin{eqnarray}
    p(\mu ; \alpha) =  \frac{\Gamma(\alpha_{0})}{ \prod_{c=1}^{C}\Gamma(\alpha_{c})}\prod_{c=1}^{C}\mu_{c}^{\alpha_{c}-1} \ , \qquad \alpha_{0} = \sum_{c=1}^{C}\alpha_{c}\,.
\end{eqnarray}
 In the above, $\Gamma(\cdot)$ is the Gamma function. We can exponentiate the output scores of the neural network, $\alpha_{c} = e^{f_{c}(x^{*},\hat{\theta})}$, where the expectation of the predictive posterior for a particular example is related to the softmax function,
\begin{eqnarray}
    \frac{\alpha_{c}}{\alpha_{0}} &=& \left[ \frac{e^{S_{c=1}}}{\sum_{c=1}^{C} e^{S_{c}}}, \dots , \frac{e^{S_{c=C}}}{\sum_{c=1}^{C} e^{S_{c}}}\right]\, ,
\end{eqnarray}
where $S_{c} = f_{c}(x^{*},\hat{\theta})$ for brevity. The entropy of the predictive mean of the categorical distribution, or the effective ``total" uncertainty, can be expressed in analytic form through the Dirichlet concentration parameters as
\begin{eqnarray}
    \mathbbH \big[\ex_{p(\mu | \mathbf{x}^{*},\hat{\theta})}[p(y|\mu)]\big] &=& - \sum_{c=1}^{C} \frac{\alpha_{c}}{\alpha_{0}} \ln \frac{\alpha_{c}}{\alpha_{0}}\ .
\end{eqnarray}
\begin{figure}
    \centering
    \includegraphics[width=0.49\linewidth]{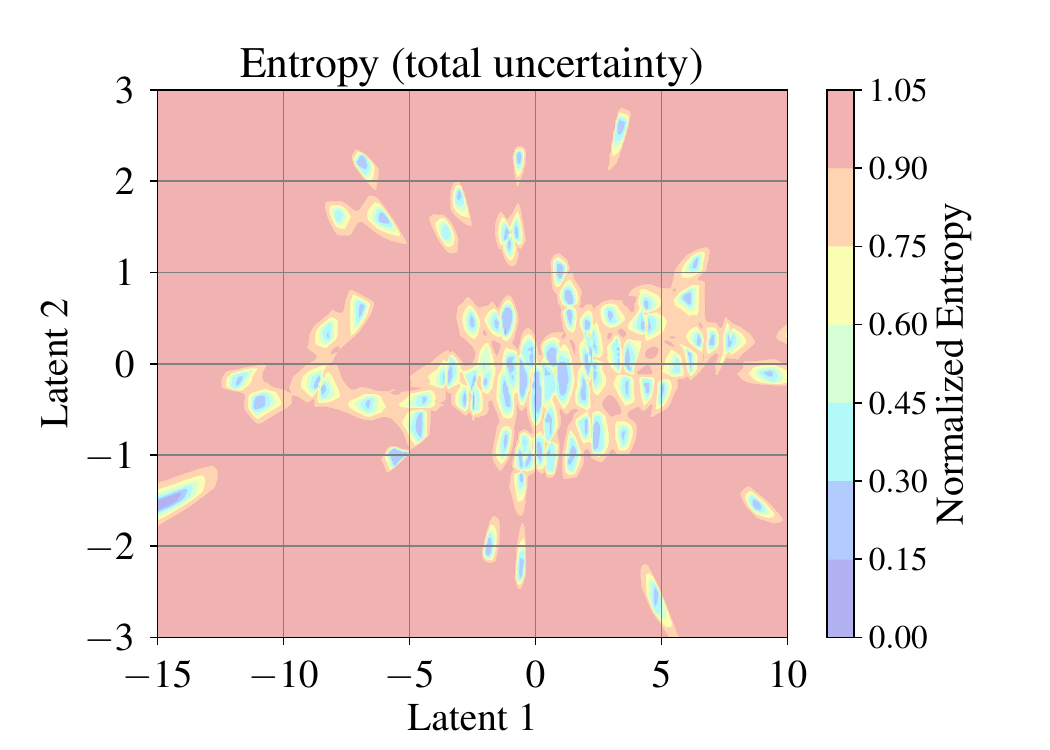}
    \includegraphics[width=0.49\linewidth]{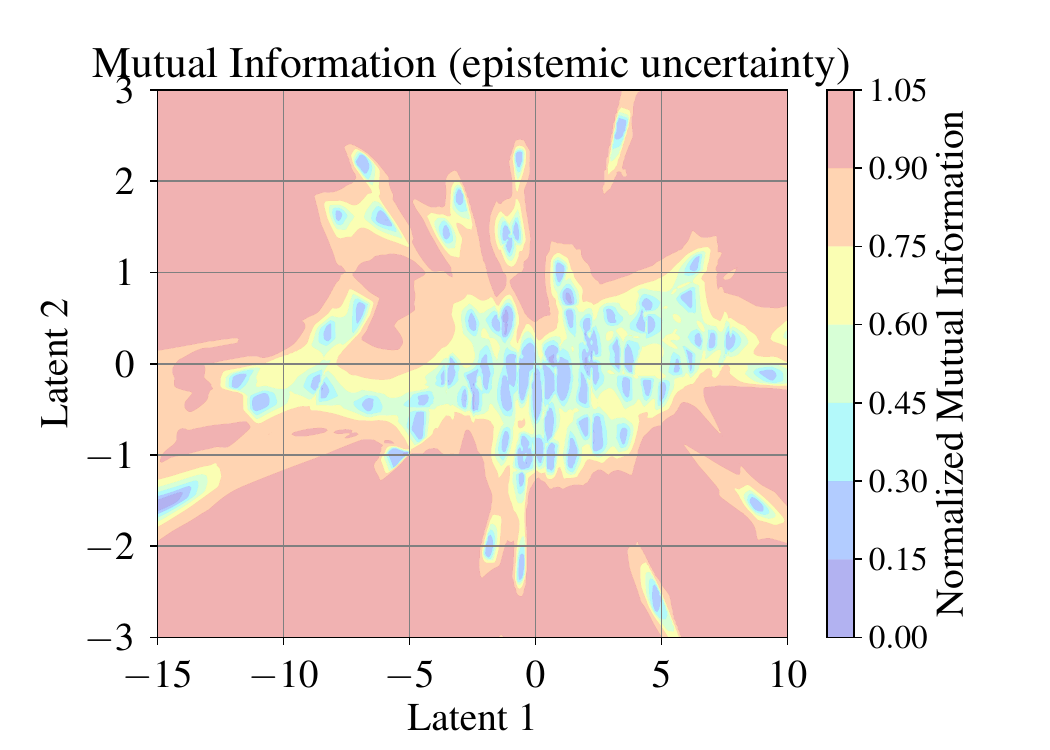}
    \caption{We demonstrate the output of the trained Dirichlet prior network (DPN) head which classifies SMEFT universes in the latent embedding space with uncertainties. In (\textit{left}) we calculate the total classification uncertainty through the entropy, and (\textit{right}) we calculate the epistemic uncertainty modeled through the mutual information demonstrating regions of OOD samples.}
    \label{fig:class_uq}
\end{figure}
In Fig.~\ref{fig:class_uq} (\textit{left}), we calculate this total entropy as a function of the coordinate in the latent space, and then normalize the entropy to the range $\mathbbH \in [0,1]$ for visualization. One can see that the Dirichlet classifier head can pull apart the SMEFT universes in the latent space, ascribing lower entropies to regions of certainty. The DPN also recognizes regions of significant overlap in the ``SM''-region indicated by higher entropies. Finally the model delineates the boundaries of the SMEFT universe with OOD samples indicated in red. These OOD samples can be further investigated through the epistemic uncertainty which is calculated using the mutual information. In Fig.~\ref{fig:class_uq} (\textit{right}), we show an example calculation of the mutual information (details given in the following section) which indicates regions where the embedding coordinates are OOD from the training data. Using this DPN head, we can proceed from deformations in the differential distributions to families of SMEFT Wilson coefficients, examining the entropy as a measure of the associated classification uncertainty.

\subsubsection*{Anomaly Detection}

The ability to distinguish anomalous distributions from standard physics scenarios is crucial for identifying BSM signals in collider phenomenology. The anomaly detection head described in the following section provides a convenient framework to extend this notion from a binary decision of in- {\it vs.}~out-of-distribution to a multi-valent quantifier of exactly how out-of-distribution a given instance is. We use the distributional uncertainty which can be described through the mutual information between target labels $y$ and categorical probabilities $\mu$:
\begin{eqnarray}
    \mathcal{I}[y,\mu | x^{*}, \mathcal{D}] &=& \mathbbH\big[\mathbbE_{p(\mu|x^{*},\mathcal{D})}[p(y|\mu)]\big]  - \mathbbE_{p(\mu | \mathbf{x}^{*},\mathcal{D})}\big[\mathbbH [p(y|\mu)]\big]\ .
\end{eqnarray}
Through the Dirichlet distribution, we can approximate this quantity using the estimated DPN concentration parameters:
\begin{eqnarray}
    \mathcal{I}[y,\mu | \mathbf{x}^{*}, \hat{\theta} ] &=& - \sum_{c=1}^{C} \frac{\alpha_{c}}{\alpha_{0}} \ln \frac{\alpha_{c}}{\alpha_{0}} + \sum_{c=1}^{C}\frac{\alpha_{c}}{\alpha_{0}}\big(\psi(\alpha_{c} + 1) - \psi(\alpha_{0} + 1) \big)\ . \nonumber
\end{eqnarray}

\begin{figure}
    \centering
    \includegraphics[width=0.32\linewidth]{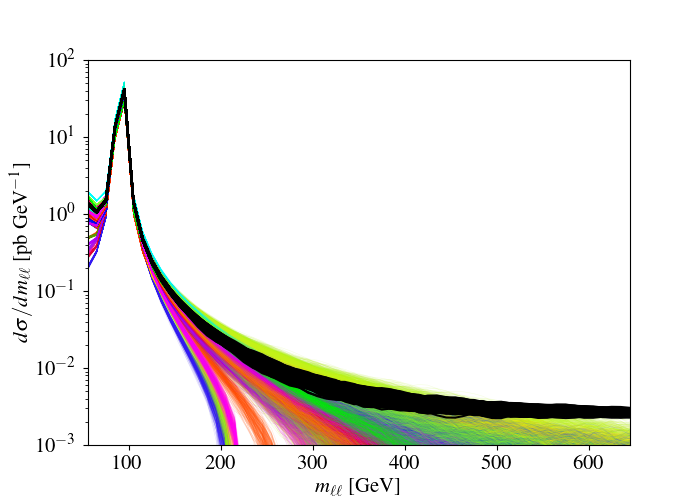}
    \includegraphics[width=0.32\linewidth]{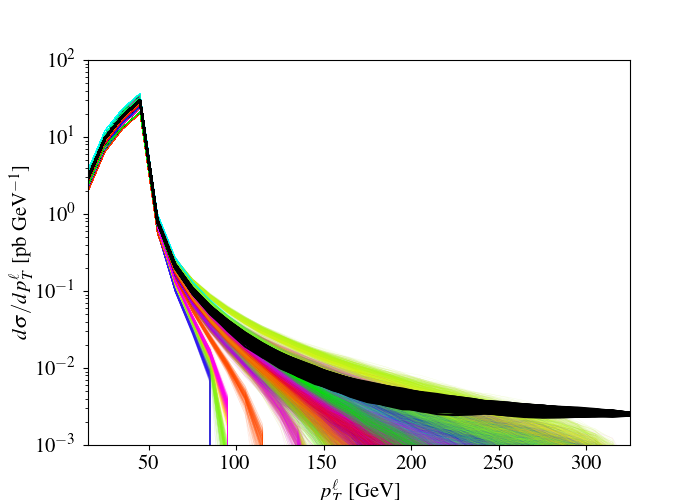}
    \includegraphics[width=0.32\linewidth]{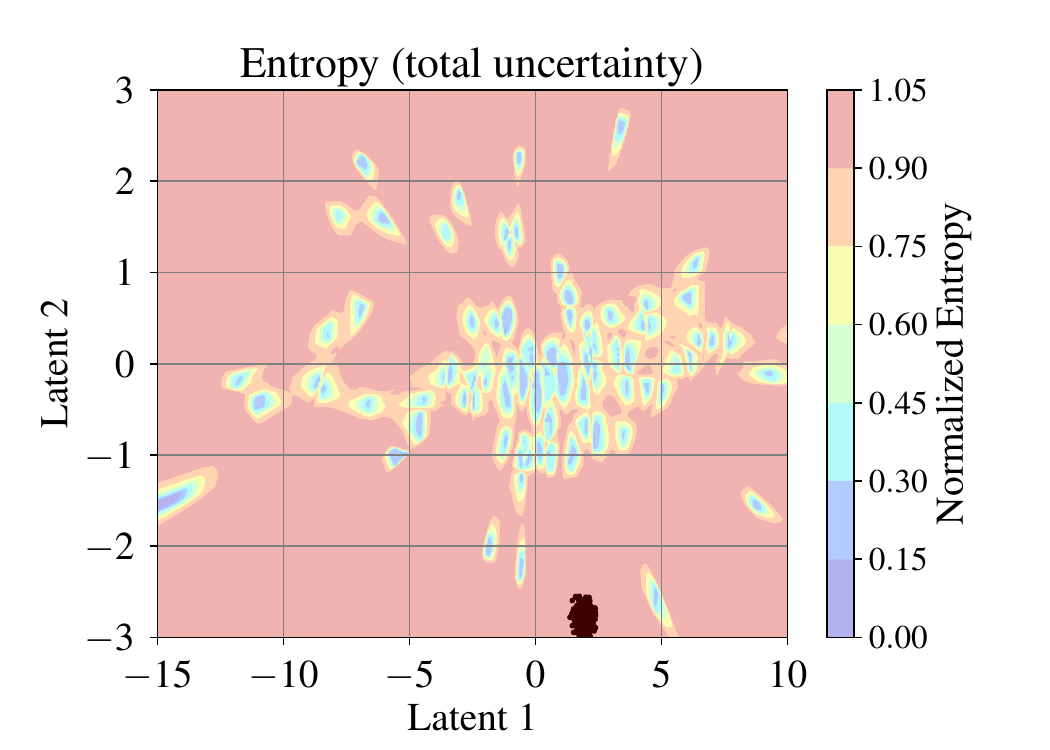}
    \includegraphics[width=0.32\linewidth]{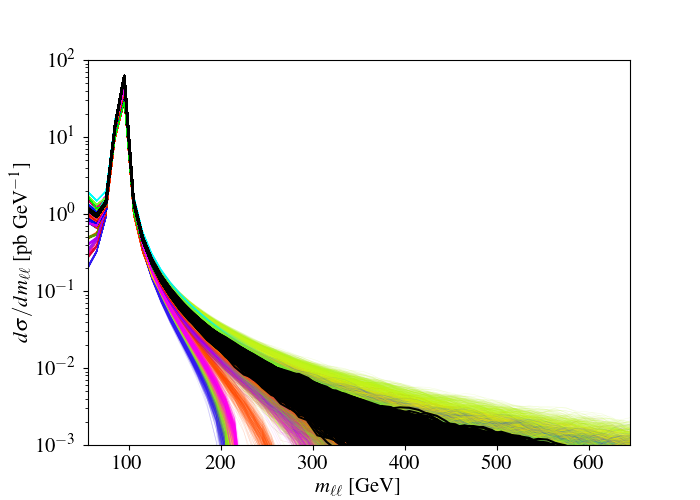}
    \includegraphics[width=0.32\linewidth]{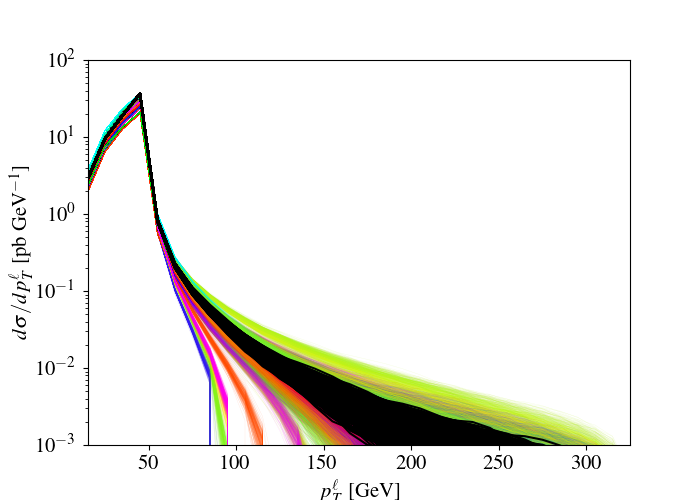}
    \includegraphics[width=0.32\linewidth]{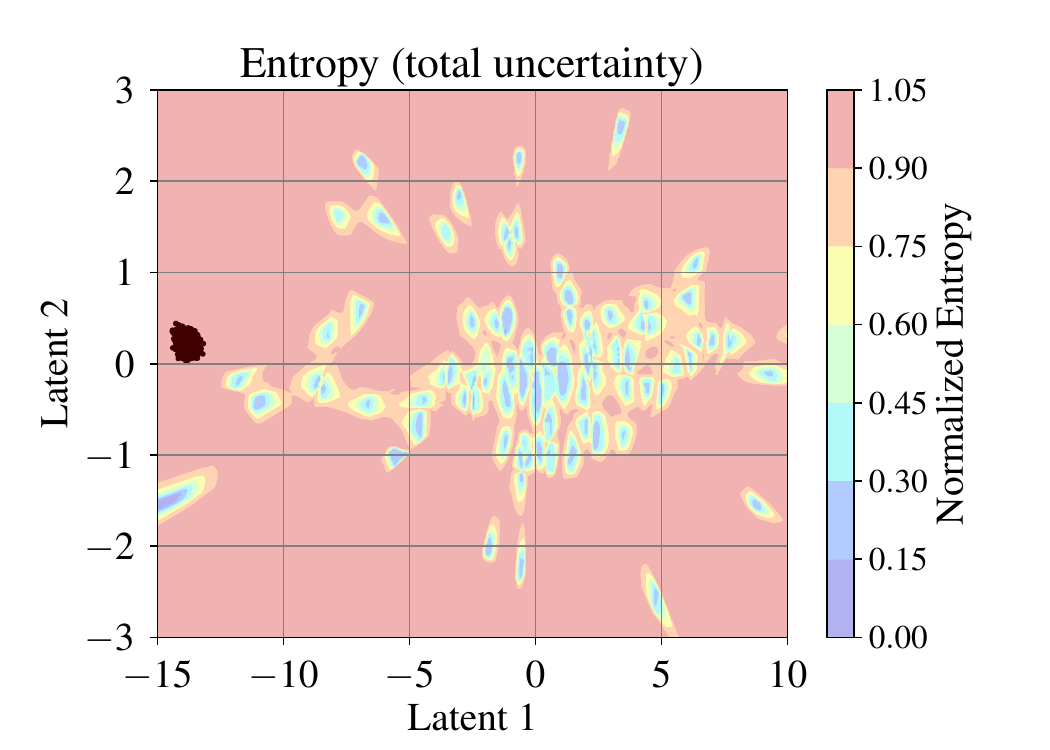}
    \includegraphics[width=0.32\linewidth]{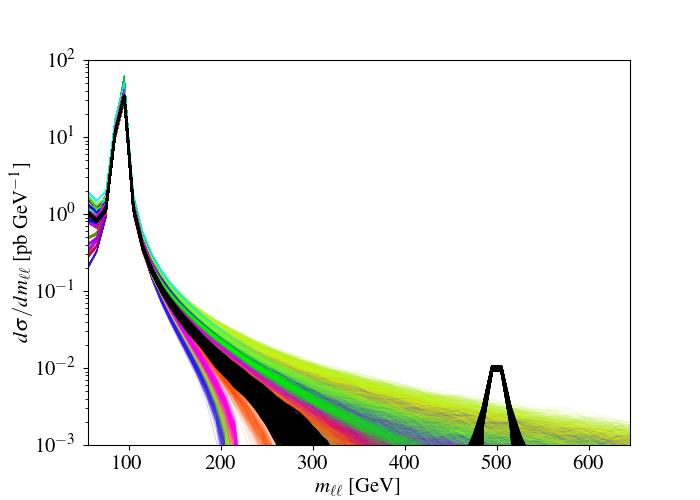}
    \includegraphics[width=0.32\linewidth]{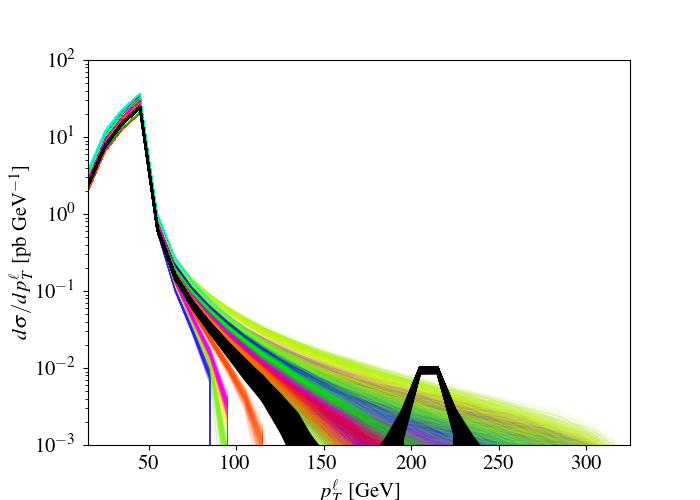}
    \includegraphics[width=0.32\linewidth]{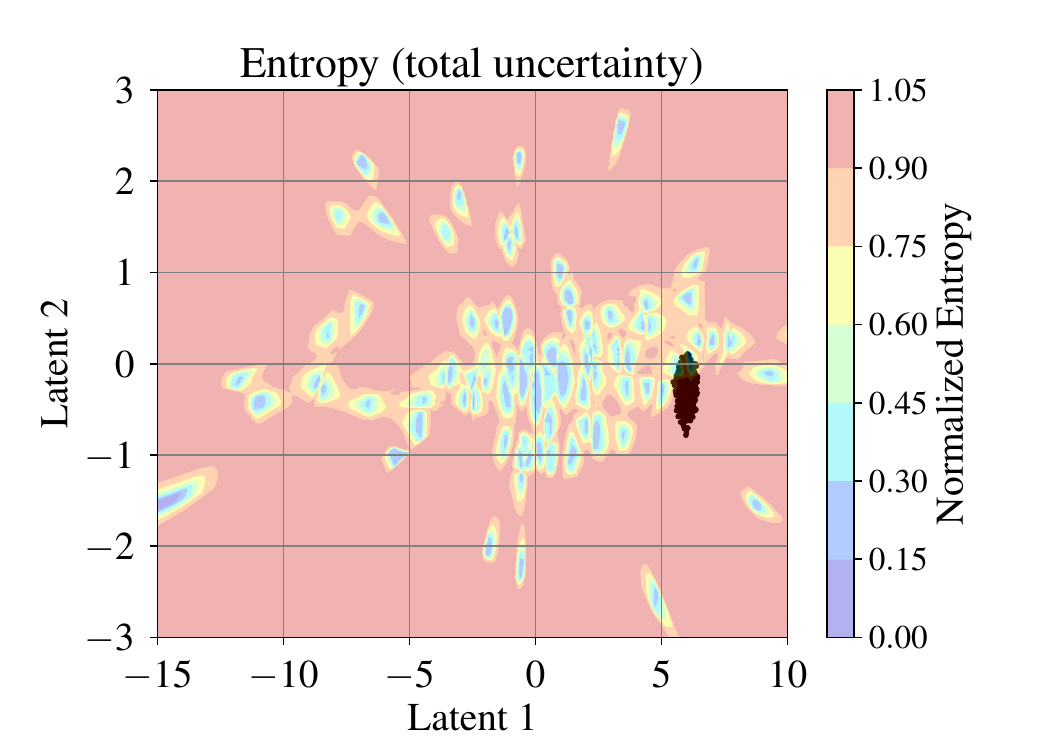}
    \includegraphics[width=0.32\linewidth]{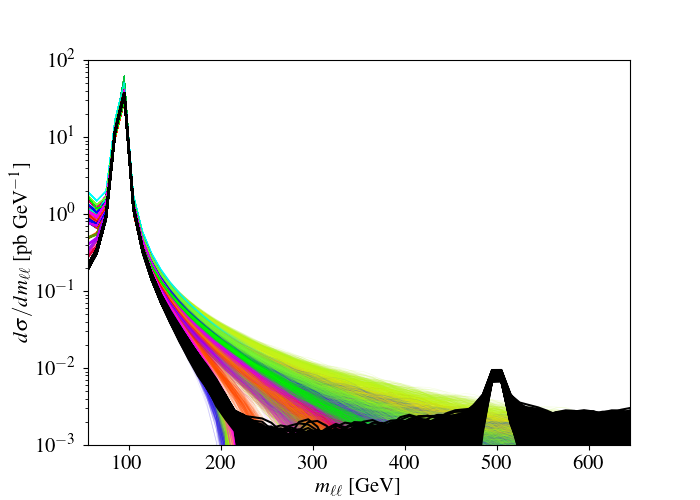}
    \includegraphics[width=0.32\linewidth]{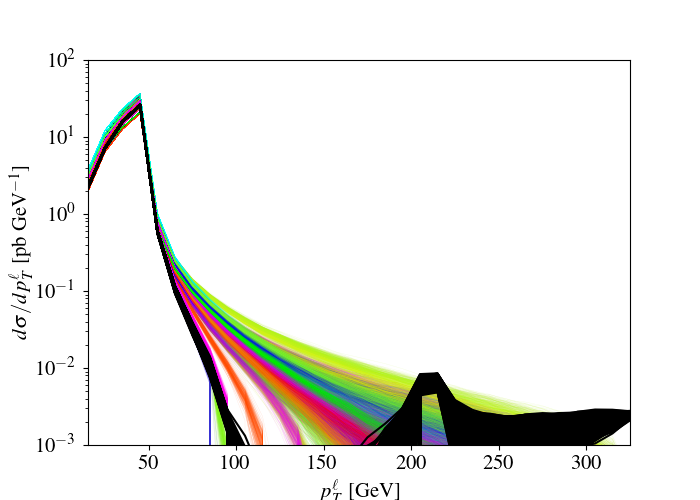}
    \includegraphics[width=0.32\linewidth]{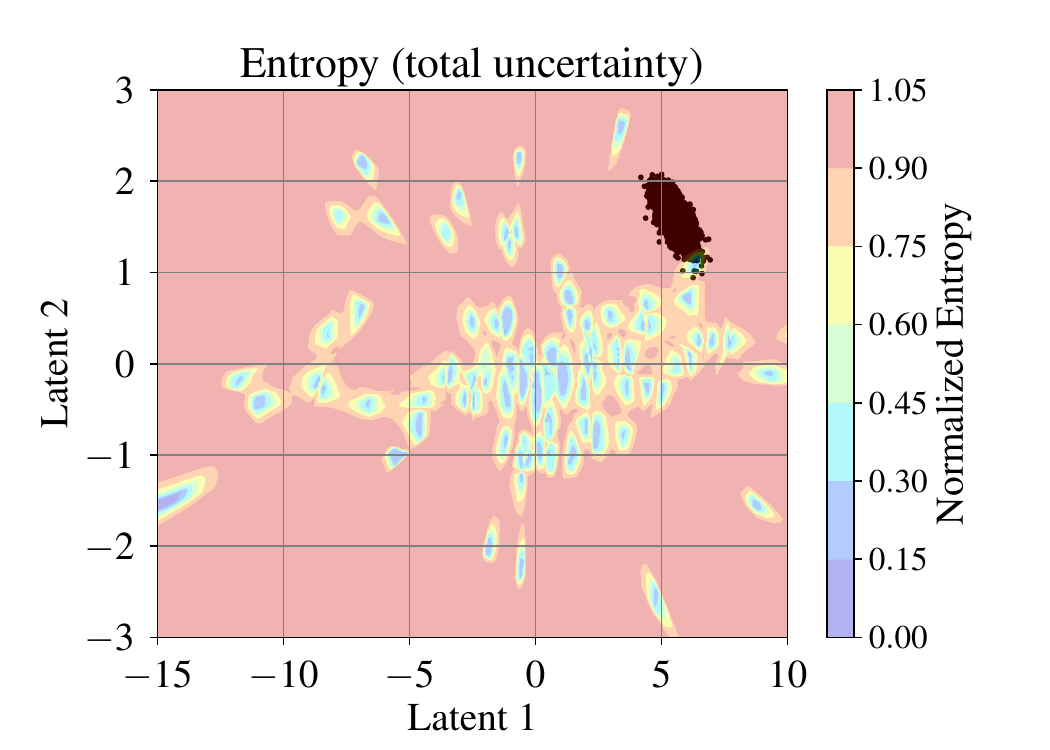}
    \caption{A case study using the anomaly detection head of the foundational embedding to investigate various contributions described in the text of anomalous New Physics that lies outside of the initial training distribution. Such examples of the differential distributions are shown as well as their embedding in the latent space with an overlay of the associated entropy.}
    \label{fig:anomaly_detection}
\end{figure}

In Fig.~\ref{fig:anomaly_detection}, we demonstrate four use cases of the anomaly detection head in identifying different types of anomalous signatures one could possibly encounter as they manifest in a differential distribution. In the \textit{(first row)} we demonstrate the effects of a non-physical excess in the tail of the invariant mass and transverse momentum distributions as shown in black in contrast to the training dataset. In the third panel of this row one can see these distributions are mapped to a region of the entropy map which has a high entropy, and therefore is anomalous compared to the input distributions. The (\textit{second row}) demonstrates a shift in the expected events around the $Z$-boson peak. The  (\textit{third row}) demonstrates an example of a Breit-Wigner peak around $500$ GeV representing the potential insertion of some New Physics at low-energy. And finally in the (\textit{third row}) we show a combination of the three New-Physics insertions which are deemed anomalous. One should notice that the embedded differential distributions are mapped in every case to regions of max entropy outside of the region of the initial training dataset --- therefore, the distributions are identified by the anomaly detection head.

\subsubsection*{Nearest-Neighbor Retrieval}

A key capability of a foundation model is the ability to pull relevant information on which the model was initially trained to respond to the user's query. We can translate this task into a physics use-case which complements phenomenological problems of interest. Given a set of uncertainties on a kinematic distribution, can we identify which SMEFT scenarios can be captured by these uncertainties. We developed a nearest-neighbor retrieval tool which embeds the user's input distribution with uncertainties and draws contours in the embedding space. All SMEFT universes within these contours are then returned.

\begin{figure}
    \centering
    \includegraphics[width=0.65\linewidth]{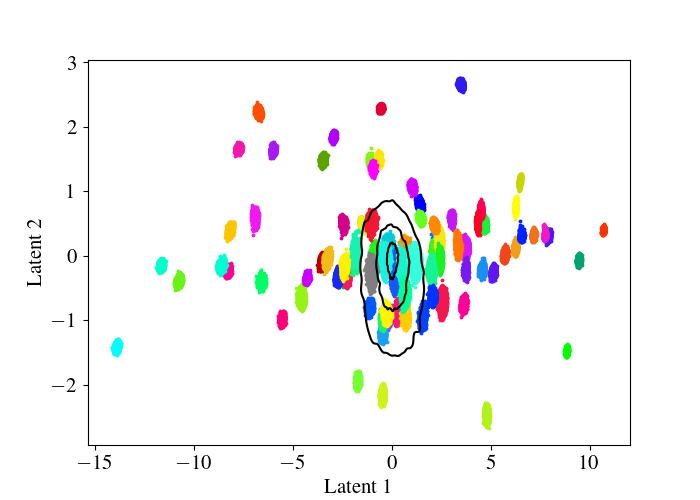}
    \caption{A demonstration of the nearest-neighbor retrieval in the foundational embedding spaced where we embed the SM distributions and show 1, 3, and 6 sigma contours.}
    \label{fig:nn_retrieval}
\end{figure}

\begin{table}[!b]
\centering
\renewcommand{\arraystretch}{1.2}
\begin{tabular}{|c|p{0.7\linewidth}|c|}
\hline
\textbf{Level} & \textbf{SMEFT universes Contained} & \textbf{Count} \\
\hline
$1\sigma$ & 5, 7, 20, 41, 53, 57, 75, 81, 82 & 9 \\
\hline
$3\sigma$ & 5, 7, 19, 20, 23, 41, 53, 57, 75, 81, 82, 97 & 12 \\
\hline
$6\sigma$ & \raggedright
5, 7, 8, 12, 19, 20, 22, 23, 26, 41, 48, 52, 53, 57, 64, 66, 71, 72, 75, 77, 81, 82, 93, 94, 97
& 25 \\
\hline
\end{tabular}
\caption{SMEFT universes which are contained within the $1\sigma$, $3\sigma$, and $6\sigma$ confidence levels of the SM calculation for neutral-current Drell--Yan.}
\label{tab:nn_table}
\end{table}

In Fig.~\ref{fig:nn_retrieval}, we embed the SM kinematical differential distributions for the neutral-current Drell--Yan process with its uncertainties as calculated from \MG and overlay the contour on our foundational embedding space. We perform this with $1\sigma$, $3\sigma$, and $6\sigma$ uncertainties on the SM calculation to capture a varying degree of consistency with the SM and the corresponding selections of SMEFT universes. One can see that the SM result sits at the center of the embedding space --- effectively, at coordinate $(0,0)$; as such, the $1\sigma$, $3\sigma$, and $6\sigma$ contours noted above form roughly concentric ellipses about this central locus. In Table~\ref{tab:nn_table}, we show the results of this nearest-neighbor retrieval head on our foundational embedding space organized by the confidence level probed and the SMEFT universes captured by the tool. At $1\sigma$ the head finds a total of 9 models which fall within the confidence region, for $3\sigma$ there are 12 models, and finally for $6\sigma$ the head captures 25 total SMEFT universes which are comparable with the SM uncertainties.

\begin{figure}
    \centering
    \includegraphics[width=\linewidth]{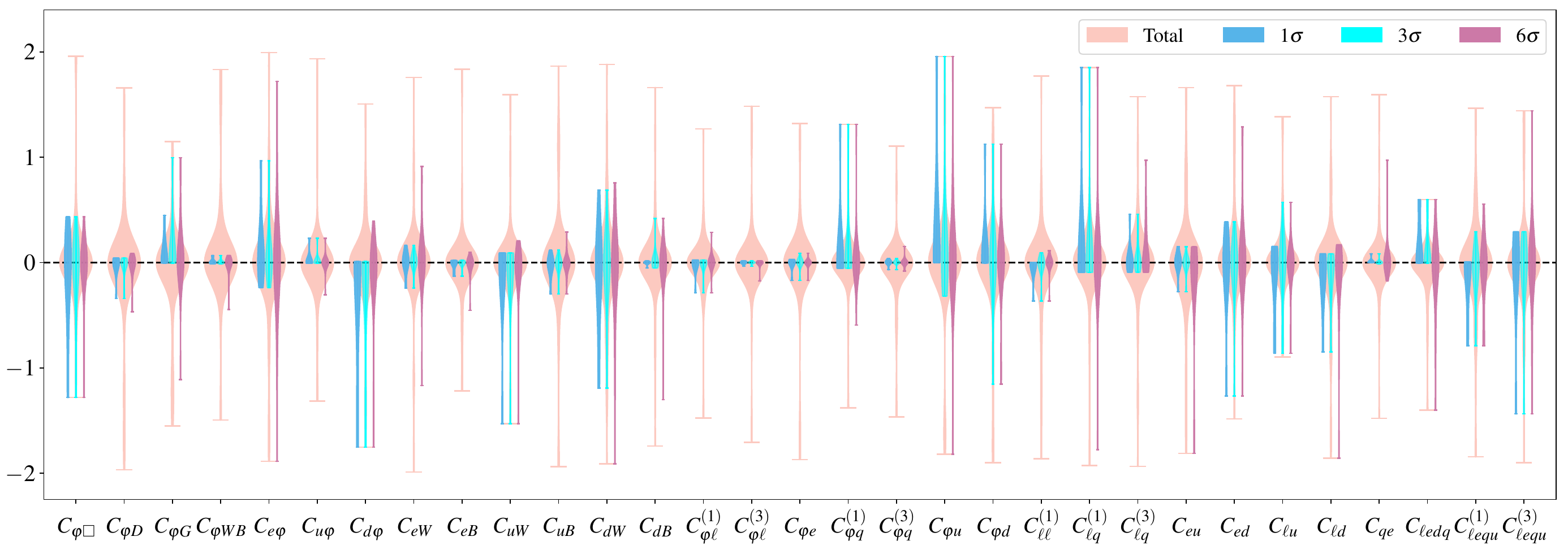}
    \caption{A comparison of the ensembled Wilson coefficients which have been captured by the nearest-neighbor retrieval head as a violin plot for each SM confidence level. This is compared to the initial training data to show potential shifts in the summary statistics.}
    \label{fig:wc_scan}
\end{figure}

We can ensemble the captured Wilson coefficients and plot a summary of the statistical distribution for each SM confidence level in violin plots, which illustrate the allowed ranges for each parameter in addition to graphical representations of the concentration of SMEFT universes within those ranges --- visualized as transverse bumps along each scan. We compare this against the distribution of the initial Wilson coefficients used in training in Fig.~\ref{fig:wc_scan}. One should note that this is not an apples-to-apples comparison with a global fit in the sense that a standard global analysis would scan the Wilson coefficient space and minimize the $\chi^{2}$; however, this result does gesture at potential connections to global analyses due to the fact that foundation models may contain the likelihood function or be fine-tuned to represent the likelihood function as a downstream task. In the current demonstrator example, the embedding space has no direct knowledge of the SMEFT parameters that enter, merely the effect of the Wilson coefficients on the differential distributions. Since the Wilson coefficient effect on the differential cross sections is characterized by degeneracies --- meaning many combinations of Wilson coefficients can lead to similar-looking deformations --- it is possible that in the embedding's current form there are significant hidden correlations between sampled Wilson coefficients that manifest as over- or under-constraints in Fig.~\ref{fig:wc_scan}. Because of this, we leave such studies investigating the downstream task of inference for future work. Preliminarily, one can see in Fig.~\ref{fig:wc_scan} that there are SMEFT Wilson coefficients which are moderately constrained with respect to the initial training data given in red. Such examples include $C_{\varphi D}, C_{\varphi WB}, C_{u\varphi}, C_{eB}$, and $C_{uB}$ as well as $C_{\varphi \ell}^{(1)}, C_{\varphi \ell}^{(3)}, C_{\varphi e}, C_{\varphi q}^{(3)}, C_{\ell \ell}^{(1)}$, across the confidence levels of the SM. We also note that examples like $C_{\varphi \Box}, C_{d\varphi}, C_{uW}, C_{dW}, C_{\varphi u}, C_{\ell q}^{(1)}, C_{ed}, C_{\ell e q u}^{(1)}, C_{\ell e q u}^{(3)}$ saturate the initial training data distribution and therefore are under-constrained within the current demonstrator model. 

\begin{figure}
    \centering
    \includegraphics[width=\linewidth]{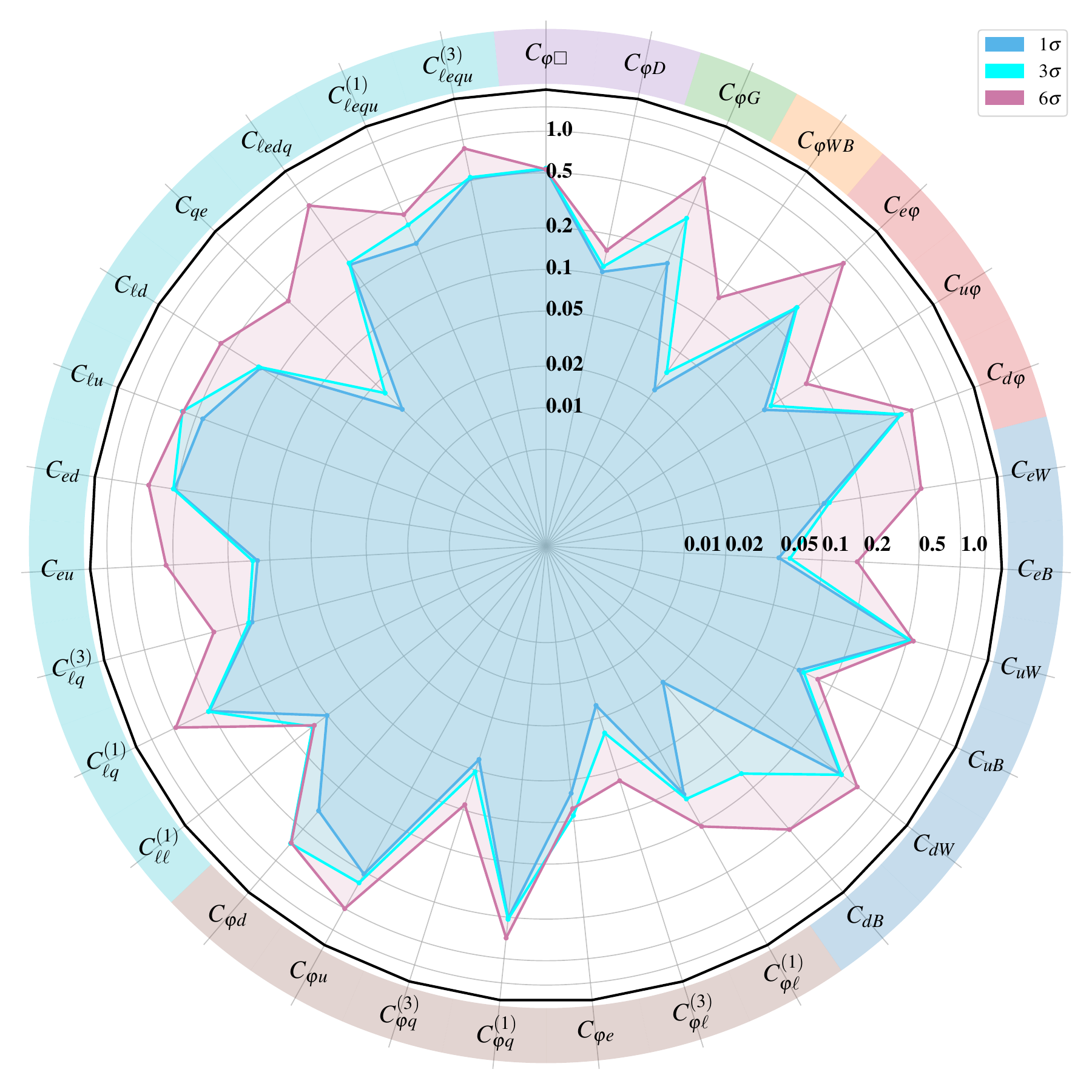}
    \caption{Constraints on the relevant dimension-6 SMEFT Wilson coefficients for the neutral-current Drell--Yan process plotted as ratios of the captured ranges from the nearest-neighbor retrieval head as compared to the initial training distributions.}
    \label{fig:wc_radial}
\end{figure}

Given the fact that significant asymmetries can be seen in the reduced ranges allowed to the Wilson coefficients in Fig.~\ref{fig:wc_scan}, it is also
instructive to examine the extent to which the 1, 3, and 6$\sigma$ consistency requirements with the SM limit the total range of each
parameter. We visualize this information as a complement to Fig.~\ref{fig:wc_scan} in the ``radial plot'' of Fig.~\ref{fig:wc_radial}. In
constructing this plot, we simply compute the absolute range allowed to each Wilson coefficient as $\Delta C_i = C^\mathrm{up}_i - C^\mathrm{down}_i > 0$,
normalizing each of these allowed ranges to the corresponding unrestricted range ({\it i.e.}, total) range used in training the foundation
model before applying the nearest-neighbor selection. The resulting visualization provides an intuitive picture of the extent to which consistency
with the SM to varying statistical levels selects or excludes particular SMEFT universes within the full trained ensemble. As a general
rule, larger excursions away from the SM (out to the 6$\sigma$ level) delineate broader subsets of the embedding space and correspond
to looser ranges of the Wilson coefficients. These more permissive selections can be identified with the outermost radial bands in Fig.~\ref{fig:wc_radial} --- that is, the purple 6$\sigma$ set of spokes. Tighter levels of consistency with the purely SM prediction meanwhile produce the correspondingly more central sets of radial bands. Notably, the particularly high degree of constraining power of the neutral-current Drell--Yan cross sections admitted to training can be seen by the rapid collapse of the radial coordinate for select coefficients: {\it e.g.}, $C_{\varphi W B}$ and $C_{qe}$, among others.

\section{Outlook and Conclusions}
\label{sec:conclusions}

In this work, we have presented the first theory-based foundation model for organizing SMEFT collider signatures in a semantic latent embedding; as an initial application of the concept, we have concentrated on the neutral-current Drell--Yan process, for which we train to a representative basis of inclusive observables at the singly differential level.
We trained a contrastive encoder on theory-level Monte Carlo replicas for the full kinematical dependence of these differential distributions as calculated within \MG; we also modeled uncertainties to construct a low-dimensional latent manifold encoding SMEFT-induced deformations away from the SM.
We interrogated the learned representation and demonstrated the interpretability of the encoded representation with respect to a range of phenomenological behaviors in the cross sections as given by the underlying SMEFT theory. Also, consistent with the foundational nature of the embedding model, we showed multiple examples of downstream tasks, including classification with uncertainty quantification, anomaly detection, and nearest-neighbor retrieval.
In the latter case, we leveraged the nearest-neighbor retrieval to illustrate the connection between cross section-level
constraints consistent with the SM and the corresponding selection or exclusion of specific SMEFT scenarios.
These demonstrations of the initial framework --- though restricted to leading-order SMEFT --- already illustrate that learned embeddings can capture a portion of the intrinsic structure of the SMEFT theory and can serve as a foundation for a scalable, multi-process study with the potential for complementing future global analyses.

The present study is a first demonstration of a trained foundational embedding based in the modern theory for collider observables. As such, it is extensible towards a more comprehensive model of SMEFT signatures in collider phenomenology.
A next step toward the realization of an enlarged foundation model is the construction of a more comprehensive training corpus that includes: multi-process, multi-energy experimental training sets which include observable with additional SMEFT constraining power beyond those considered here. These include charged-current Drell--Yan, diboson, top, Higgs, DIS, and jet measurements; an expanded list of theoretical observables including rapidity, angular spectra, and double-differential distributions; and finally a physics-informed sampling algorithm for efficient SMEFT universe generation.
To match current phenomenological standards, higher-order corrections, including NLO in both QCD and SMEFT, and the $\mathcal{O}(\Lambda^{-4})$ contributions from dimension-6 squared and linear dimension-8 terms, will be important. In addition to these ingredients, it will also be necessary to explore a fuller treatment of potential SM uncertainties like the PDF parameterization dependence and variations in the renormalization- and factorization-scales ($\mu_{R,F}$); similarly, examining the interplay with systematic effects, especially once experimental data are introduced, will be a vital consideration.

As the above training corpus grows to include a broader array of predictions based upon the underlying SMEFT, machine learning-based developments must compensate the challenges associated with this enlarged dataset. Future work will likely require transformer architectures to handle the larger data curation, as well as an intentional tokenization scheme to incorporate as much meta-data as possible. A multi-objective pretraining scheme will be designed to enrich the semantic structure of the latent space, such as perturbative-order classification, SMEFT truncation identification, PDF--vs--BSM disentangling, and masked-bin reconstruction.
A systematic study of interpretability, including the alignment of latent directions with 
operator classes, the response to controlled Wilson-coefficient perturbations, and the geometry of RG flows, will also be critical for establishing the physical fidelity of this advanced learned representation.

Together, the developments in this study point towards a more comprehensive theory-based foundation model whose latent embedding(s) have the capability to mediate a diverse set of processes and datasets, culminating in a class of downstream analyses. Realizing such a program would complement traditional SMEFT collider phenomenology, while also enfolding an array of global fits into a geometry-based framework for exploring the underlying SMEFT theory for current and future collider facilities.

%
%

\section{Acknowledgments}
We thank B.~Assi, S.~Dawson, and A.~Martin for useful discussions at several stages
during the completion of this study.
This work at Argonne National Laboratory was supported by the U.S.~Department of Energy under contract DE-AC02-06CH11357.
We gratefully acknowledge use of the Bebop supercomputer in the Laboratory Computing Resource Center at Argonne National Laboratory for the calculations shown in this study.
%
%

\bibliographystyle{JHEP}
\bibliography{references} 

\end{document}